%% file: Main.tex
\documentclass[lettersize,journal]{IEEEtran}

\usepackage{soul}
\usepackage[utf8]{inputenc}
\usepackage{amsthm}

\usepackage{epsfig, subfigure, amsmath, amssymb}
\usepackage{textcomp}
\usepackage{url}
\usepackage{wrapfig,lipsum,booktabs}
\usepackage{cite}
\usepackage{multirow}
\usepackage{amsmath,amssymb,amsfonts}
\usepackage[ruled,vlined,linesnumbered]{algorithm2e}
\usepackage{algorithmic}
\usepackage{graphicx}
\usepackage{textcomp}
\usepackage{multirow}
\usepackage{xcolor}
\usepackage{amsmath}
\usepackage{makecell}
\usepackage{paralist}

\newcommand{\rev}[1]{\textcolor{black}{#1}}   
\newcommand{\revb}{\color{black}}              
\begin{document}

\title{Attack-Resiliency Analytics for Wide-Area Control Systems in Smart Grids \hfill} 


\author{{Mohammad Zakaria Haider, \IEEEmembership{ Member, IEEE}, Prabin~Mali, \IEEEmembership{ Member, IEEE}, Nur~Imtiazul~Haque, \IEEEmembership{ Member, IEEE},\\ Muhammad~Nadeem, \IEEEmembership{Member, IEEE}, Sumit~Paudyal,~\IEEEmembership{Senior Member,~IEEE}, Mohammad Ashiqur Rahman, \IEEEmembership{Senior Member, IEEE}}

\thanks{This work is supported by the Department of Energy (DOE) under Award\# DE-CR0000024. Any opinions, findings, conclusions, or recommendations expressed in this material are those of the authors and do not necessarily reflect the views of the DOE or the U.S. Government.
    
    Mohammad Zakaria Haider is a Ph.D. student in the Department of Electrical and Computer Engineering (ECE) at Florida International University (FIU) and a member of FIU's Analytics for Cyber Defense (ACyD) Lab, Miami, FL, USA (email: mhaid010@fiu.edu ).
    
    Prabin Mali is a Ph.D. student in the Department of Electrical and Computer Engineering (ECE) at Florida International University (FIU) and a member of FIU's Power System Computational Laboratory, Miami, FL, USA (email:pmali004@fiu.edu).

    Nur Imtiazul Haque is an Assistant Professor in the Department of Computer Science at Northern Illinois University, Illinois, USA (email: nhaque@niu.edu).
    
    Muhammad Nadeem is a Postdoctoral Associate at Knight Foundation School of Computing and Information Sciences (KFSCIS) at FIU and a member of the ACyD Lab, Miami, FL, USA (email: mnadeem@fiu.edu).
    
    Sumit Paudyal is a Professor of the ECE Department at Florida International University (FIU) and the Director of the Power System Computational Laboratory, Miami, FL, USA (email: spaudyal@fiu.edu).
    

    Mohammad Ashiqur Rahman is an Associate Professor of the KFSCIS at FIU and the Director of the ACyD Lab, Miami, FL, USA (email: marahman@fiu.edu). 
   	}
}


\maketitle

\thispagestyle{plain}
\pagestyle{plain}

\begin{abstract}
\input{sections/Abstract}
\end{abstract}


\input{sections/Introduction}

\input{sections/Nomenclature}
\input{sections/Technical_Overview}
\input{sections/Problem_Formulations}

\input{sections/Evaluation}

\input{sections/Discussion}
\input{sections/Conclusion}

\bibliographystyle{ieeetr}
\bibliography{sections/References}

\end{document}

%% file: sections/Abstract.tex
Wide-area monitoring, protection, and control (WAMPAC) systems damp inter-area oscillations in large interconnected grids, but their reliance on synchronized PMU measurements carried over wide-area networks exposes them to false data injection (FDI) attacks. This paper presents an attack-resiliency analytics framework that formally models the coupled dynamics of the wide-area damping loop, automatic generation control, and the governor and excitation systems, and formulates the optimal stealthy FDI attack as a mixed-integer linear program (MILP). The anomaly detection model (ADM) enters as a replaceable constraint set: either a static bad-data detection (BDD) rule or a boundary learned from benign operation calibrated to a common false-positive rate. On the IEEE 39 and 118 bus systems, with reference dynamics validated on an OPAL-RT hardware-in-the-loop testbed, the optimal attack with wide-area access attains 3.3 and 8.1 times the benign objective and reaches a 0.5~Hz frequency excursion more than twice as fast as an attack confined to automatic generation control. Learned boundaries reduce the attack objective by 15.5–55.0\% and prevent over-frequency relay trips in all configurations, with feasible stealthy attacks remaining in every case, and residual risk tracks the width of the learned boundary rather than the detector family.


%% file: sections/Introduction.tex
\section{Introduction}
\IEEEPARstart{M}odern power systems are extensive, interconnected networks whose stability depends on maintaining frequency, rotor angle, and voltage within narrow operating bands. Small disturbances can excite low-frequency oscillations (LFOs) in the 0.1-2\, Hz range, classified as local (0.8-2\, Hz) or inter-area (0.1-0.8\, Hz)~\cite{Jamsheed20PESGRE}. If not adequately damped, these oscillations can cascade into blackouts, as documented in the Hydro-Qu\'{e}bec (0.6\, Hz LFO)~\cite{Kamwa00_blackout_2}, Brazil North-South interconnection (0.15\, Hz)~\cite{Martins99_blackout_3} events, while the 2012 India blackout, triggered by transmission-line overloading, shows how a local stability failure can cascade across interconnected regions~\cite{Alhelou2019ASO_blackouts_4}. Local controllers such as power system stabilizers (PSS) and automatic voltage regulators (AVR) can dampen local-mode LFOs, but inter-area oscillations (IAO) require wide-area coordination~\cite{Sarkar20JAS}.

Wide-area monitoring, protection, and control (WAMPAC) systems address this gap by leveraging synchronized phasor measurement unit (PMU) data, transmitted via phasor data concentrators (PDCs) using protocols such as IEEE~C37.118, to estimate system states and generate coordinated control actions in real time~\cite{Xin15, Chakrabortty10ISGT}. While WAMPAC substantially improves oscillation damping compared to SCADA-based control~\cite{KrommydasISGT22}, its reliance on networked PMU telemetry introduces a broad attack surface that localized defenses cannot adequately cover. False data injection (FDI) attacks can compromise state estimation by manipulating frequency, voltage, and power flow measurements to evade anomaly detection models (ADMs), with optimized attack vectors posing a particularly severe threat as grid complexity increases~\cite{Vahidi23ICST, Wlazlo21IET}.

Existing work on FDI attacks against power systems has primarily focused on isolated control loops. Studies such as~\cite{Rui16ICCPS, Jafari23TSG, jafari23tifs, haider2024PESGM} analyze attacks on AGC and load-frequency control (LFC), demonstrating frequency instability and relay tripping, while~\cite{ref_fdi_agc_rl} explores attacks on AVR and~\cite{Rajabi2021} on synchrophasor-based mode estimation. However, these analyses treat controllers independently and overlook the coupled dynamics among AGC, AVR, PSS, and WAMPAC. Furthermore, most prior work evaluates attacks against static bad-data detection (BDD) rules, which struggle to detect sophisticated, gradually escalating attack vectors. A holistic framework that jointly models the interdependencies across all control tiers and evaluates attacks against both static and ML-based detection is essential for a comprehensive grid security assessment. This paper develops such a framework and evaluates it on two test systems of markedly different scales. The core contributions of this work are:

\begin{itemize}
    \item We derive a formal model capturing the interdependencies among the voltage and frequency controllers of a WAMPAC-enabled grid: the proportional–integral–derivative (PID)-based wide-area damping loop, the AGC based secondary layer, the governor and excitation dynamics, and the generator swing equations. This model supplies the constraint set for an attack optimization in which the adversary maximizes the cumulative frequency and voltage excursion from nominal, subject to the coupled dynamics, operational limits, an accessibility budget over reachable controllers, and anomaly detection model (ADM)-consistency bounds.

    \item Alongside the conventional static BDD rule set, we derive per-controller stealth bounds directly from the benign PMU stream, using a family of anomaly detectors calibrated to a common benign false-positive rate, and solve the same attack problem once per detector. This turns a single attack demonstration into a controlled comparison of what different decision boundaries concede to an optimal adversary.

    \item We quantify the severity an optimal attacker attains once the wide-area channel is within reach, show that coordinated attacks reach a 0.5~Hz relay-relevant excursion more than twice as fast as an AGC-only campaign, and establish that residual severity under a learned detector is governed by the width of its decision boundary rather than by the model class producing it. The orderings hold across the 39-bus and 118-bus systems, with reference dynamics validated against an OPAL RT hardware-in-the-loop testbed.
\end{itemize}

The remainder of this paper is organized as follows. Section~\ref{sec:related_works} reviews the related literature. Section~\ref{sec:technical} develops the mathematical model of WAMPAC. Section~\ref{sec:problem_formulation} formalizes the optimal FDI attack problem. Section~\ref{sec:evaluation} presents the testbed setup, test cases, and evaluation results. Section~\ref{sec:discussion} discusses insights, limitations, and future directions. Section~\ref{sec:conclusion} concludes the paper.

\section{Related Work}
\label{sec:related_works}

The energy sector was the fourth-most-attacked industry in 2022, accounting for 10.7\% of all cyberattacks~\cite{jenifer}. High-profile incidents such as Triton/Trisis~\cite{ref_12}, Stuxnet~\cite{stuxnet}, and Dragonfly~\cite{Dragonfly} demonstrate that adversaries with access to real-time control center data can cause physical damage to critical infrastructure. WAMPAC systems, which play a vital role in damping oscillations through real-time monitoring and control~\cite{Segundo21WAC, Zhao23ISGT}, are particularly exposed because their distributed architecture relies on PMU data transmitted over communication links using protocols such as C37.118~\cite{Vahidi23ICST}, DNP3~\cite{Wlazlo21IET}, and Modbus~\cite{HUITSING200837}. By deceiving PDCs into connecting with spoofed PMUs, attackers can infiltrate wide-area networks and disrupt power system stability~\cite{Swain24Access, Sadreazami18SIPN, Pasqualetti}.

WAMPAC systems are critical for damping inter-area oscillations by measuring phase angle differences and power flows between zones and triggering coordinated control actions through PSS and FACTS devices~\cite{Jamsheed22PESGRE, Ranjbar17ICEE, Benayad20ENERGYCon}. FDI attacks can undermine this capability by manipulating the PMU measurements that drive the wide-area control signal, thereby impairing oscillation detection and mitigation~\cite{Jafari22IAS, Jafari23TSG}. Prior FDI attack studies can be broadly categorized by their scope. Works such as~\cite{Jafari23TSG, jafari23tifs} formulate optimal attacks against AGC and LFC in isolation, demonstrating their potential to cause frequency instability and relay tripping. Tan et al.~\cite{Rui16ICCPS} proposed an optimal FDI sequence to rapidly breach safety thresholds, while~\cite{Wu18ITC} framed the problem as a general numerical optimization challenge. However, these studies analyze individual control loops without accounting for the cross-coupling effects between frequency and voltage controllers. Random attacks on AGC and AVR~\cite{Rajabi2021, Vahidi23ICST} produce high-magnitude signals that are readily detectable by ADMs, making low-magnitude optimal attacks the more realistic threat model.

Modern control centers increasingly deploy ML-based ADMs, so robustness must be assessed against attack vectors that bypass them. Adversarial ML, RL~\cite{ref_fdi_agc_rl}, and formal approaches either do not yield optimal, verifiable attacks or do not capture interdependent CPS dynamics. Formal attack-analytics frameworks such as SHChecker~\cite{haque2021novel} and SHATTER~\cite{haque2023shatter} target other domains and treat controller dynamics independently. None of these jointly models AVR, PSS, AGC, and WAMPAC, or compares static and learned detectors under one attack formulation, which is the gap this work addresses.

%% file: sections/Technical_Overview.tex
\section{TECHNICAL OVERVIEW Of WAMPAC}
\label{sec:technical}

\begin{table}[t]
\caption{Nomenclature for WAMPAC Model}
\small
\begin{tabular}{p{1.2cm} p{1.3cm} p{4.9cm}}
\hline
\textbf{Type} & \textbf{Symbol} & \textbf{Description} \\ \hline
\multirow{8}{*}{\textbf{Sets}}
& $\mathcal{B}$ & All buses \\
& $\mathcal{B}^{PV}$ & Generator buses \\
& $\mathcal{B}^{GV}$ & Generator buses with governors \\
& $\mathcal{G}, \rev{\mathcal{G}_i}$ & All generators\rev{; generators of area $i$} \\
& $\mathcal{Z}$ & All control areas \\
& $\mathcal{T}^{P,S,T}$ & Timeslots for primary/secondary/tertiary response \\
& $b,c$\rev{, $g$} & Bus\rev{/generator} indices \\
& $i,j$ & Area indices \\ \hline
\multirow{3}{*}{\textbf{Bus}}
& $\delta$ & Phase / rotor angle \\
& $\mathcal{P}_L$ & Load active power \\
& $\mathcal{V}_t$ & Bus voltage magnitude \\ \hline
\multirow{5}{*}{\textbf{Generator}}
& $\mathcal{P}_g, \mathcal{P}_m$ & Electrical / mechanical power \\
& $\omega$ & Angular frequency \\
& $\mathcal{V}_{ref}, \mathcal{V}_{pss}$ & AVR reference / PSS output \\
& $\mathcal{V}_{wac}$ & WAMPAC output signal \\
& $\mathcal{E}_{fd}$ & \rev{Field voltage} \\ \hline
\multirow{6}{*}{\textbf{Param.}}
& $\mathcal{R}$ & Governor droop \\
& $\Delta T$ & Discretization time step \\
& $\mathcal{H}$ & Generator inertia constant \\
& $\mathcal{B}_A, \mathcal{G}_A$ & Line susceptance / conductance \\
& $\mathcal{K}_{P,I,D}$ & PID gains \\ & \rev{$D$} & \rev{Load-damping coefficient} \\ \hline
\multirow{4}{*}{\textbf{ADM}}
& $\tau_\omega, \tau_v, \tau_{wac}$ & BDD thresholds (freq., voltage, WAC) \\
& $\epsilon_\omega, \epsilon_v, \epsilon_{wac}$ & Slack variables for ADM bypass \\
& $\rev{\bar{\tau}_\omega}, \bar{\tau}_v$ & Over/under frequency and voltage limits \\
& \rev{$\tau_\chi^{\mathit{ML}}, \alpha$} & \rev{Learned bound of channel $\chi$; benign FPR} \\ \hline
\multirow{3}{*}{\textbf{\rev{Attack}}}
& \rev{$\tilde{(\cdot)}, \bar{(\cdot)}$} & \rev{Injected value; compromised measurement} \\
& \rev{$\mathbb{A}^{\chi}$} & \rev{Accessibility vector of channel $\chi$} \\
& \rev{$\beta_{agc}, \beta_{wac}$} & \rev{Accessibility budgets} \\ \hline
\end{tabular}
\label{tab:nomenclature}
\vspace{-10pt}
\end{table}

\subsection{System Dynamics}
\label{sec: system_dynamics}

WAMPAC utilizes PMU data to provide real-time, synchronized grid awareness, employing state estimation to estimate generator states. Operating in a closed loop, WAMPAC mitigates inter-area oscillations (IAO) by generating a global control signal (GCS) that integrates with local PSS. Modeling the interconnected grid as virtual generators, the WAMPAC employs a PID controller driven by inter-area speed deviations:
\begin{equation}
\label{eq:wac_speed_dev}
\small \Delta \omega_{Area}^{ij} = \frac{1}{|\mathcal{G}_i|}\sum_{g \in \mathcal{G}_i} \Delta \omega^g-\frac{1}{|\mathcal{G}_j|}\sum_{g \in \mathcal{G}_j} \Delta \omega^g
\end{equation}
\begin{equation}
\label{eq:wac_signal}
\small \mathcal{V}_{wac}^{b} = \mathcal{K}_{p\text{-}wac}^b \Delta \omega_{Area}^{ij} + \mathcal{K}_{i\text{-}wac}^b \int \!\Delta \omega_{Area}^{ij} \, dt + \mathcal{K}_{d\text{-}wac}^b \frac{d(\Delta \omega_{Area}^{ij})}{dt}
\end{equation}



This wide-area signal supplements the standard PSS output ($\mathcal{V}_{pss}^b = \mathcal{K}_{pss}^b*\Delta \omega^b$) to directly modify the generator's excitation system:

\begin{equation}
\label{eq:pss_total}
\mathcal{V}_{pss\text{-}total}^b = \mathcal{V}_{pss}^b + \mathcal{V}_{wac}^{b}
\end{equation}


The AVR uses this total signal to dynamically adjust the terminal voltage $\mathcal{V}_t^b$, thereby maintaining rotor angle stability amid power-flow variations. The closed-loop error of the AVR responds directly to this modified reference:

\begin{equation}
\label{eq:avr_error}
\mathcal{V}_{e}^b= \mathcal{V}_{ref}^b - \mathcal{V}_t^b+\mathcal{V}_{pss\text{-}total}^b
\end{equation}


Concurrently, secondary control is provided by the Automatic Generation Control (AGC), which adjusts system generation to maintain a 60 Hz grid frequency and scheduled tie-line flows. The AGC determines the mechanical power reference ($\mathcal{P}_{ref}^b$) by integrating the Area Control Error (ACE):

\begin{equation}
\label{eq:ace}
\small \mathcal{ACE}^i = \Delta \omega^b \left(\frac{1}{R^b}+D^b\right) +\sum \Delta \mathcal{P}_{tie}^{ij}
\end{equation}
\begin{equation}
\label{eq:pref}
\small \mathcal{P}_{ref}^b=\mathcal{P}_g^b-\int \mathcal{K}^b \cdot \mathcal{ACE}^i \, dt
\end{equation}

A standard TGOV1 steam governor model\rev{, with time constant $T^b$,} maps this reference to the actual mechanical power $\mathcal{P}_m^b$ supplied to the generator:
\begin{equation}
\label{eq:governor}
\small \mathcal{P}_m^b = \frac{1}{T^b}\int \left( \mathcal{P}_{ref}^b{-\frac{\Delta \omega^b}{R^b}}-\mathcal{P}_m^b \right) dt
\end{equation}

Ultimately, the overall dynamic response and frequency behavior of the synchronous generators are governed by the classic swing equations. These equations couple the mechanical input ($\mathcal{P}_m^b$), electrical output ($\mathcal{P}_g^b$), and tie-line variations ($\Delta \mathcal{P}_{tie}^i$):
\begin{equation}
\label{eq:phase_angle}
\small \frac{d\delta^b}{dt} = \Delta \omega^b
\end{equation}
\begin{equation}
\label{eq:swing}
\small \frac{d \omega^b}{dt} = \frac{1}{2\mathcal{H}^b}(\mathcal{P}_m^b -\mathcal{P}_g^b -\Delta \mathcal{P}_{tie}^i- \Delta \mathcal{P}_l^i)
\end{equation}

Under normal operation, the WAMPAC continues to route corrective signals through these integrated loops until frequency and angle deviations across all adjacent areas are completely neutralized ($\Delta \delta = 0$, $\Delta \omega = 0$, $\Delta \delta_{Area_i-Area_j}=0$, and $\Delta \omega_{Area_i-Area_j}=0$).

\subsection{Protection Relays}
Under-frequency (UF), under-voltage (UV), over-frequency (OF), and over-voltage (OV) relays are essential for smart grid stability, activating when frequency or voltage deviates beyond safe thresholds to isolate faults and prevent blackouts. Per ANSI C84.1, allowable voltage deviations are $\pm 10\%$ of nominal. For frequency (nominally 60 Hz), governors correct minor deviations ($\pm 0.02$ to $\pm 0.05$ Hz). \rev{Following~\cite{Rui17TIFS}, we define} the safe frequency deviation range as $[-0.5, 0.5]$~Hz; any deviation $\Delta \omega^b$ outside this range triggers the relays.

\subsection{Anomaly Detection Model}
FDI attacks can stealthily manipulate measurements to actuate protection relays, disconnecting critical loads or generators~\cite{jafari23tifs, Jafari22IAS, Jafari23TSG}. Let $\mathbf{z}$ and $\mathbf{x}$ denote the grid measurements (e.g., $\mathcal{V}, f$) and states (e.g., $\omega, \delta$), related by $\mathbf{z} = \mathbf{H}\mathbf{x} + \mathbf{e}$, where $\mathbf{H}$ is the measurement matrix and $\mathbf{e}$ is noise. The state estimator computes $\hat{\mathbf{x}} = (\mathbf{H}^{\top}\mathbf{W}\mathbf{H})^{-1}\rev{\mathbf{H}^{\top}}\mathbf{W}\mathbf{z}$, and traditional bad data detection (BDD) raises an alarm \rev{when the residual $r=\|\mathbf{z} - \mathbf{H}\hat{\mathbf{x}}\|$ exceeds a threshold $\tau$}~\cite{Yan21SGC}. Because stealthy FDI attacks can exploit static BDD rules to push system frequency into hazardous zones undetected, we \rev{also consider detectors} that learn acceptable deviation patterns from \rev{benign} data rather than relying solely on static thresholds. To make the attack analytics robust to the choice of detector, the framework
treats the ADM as a replaceable component and evaluates the same attack problem
against four data-driven boundaries alongside the static rule.

The four detectors represent model families that a control center could realistically deploy and are trained on the same feature: the benign per-cycle rate of change of each monitored channel. DBSCAN bounds the dense benign region using a nearest-neighbor distance quantile; isolation forest identifies anomalies through random partition depth; and one-class SVM encloses the benign samples within a kernel decision surface. The LSTM performs one-step-ahead prediction and flags ramps whose prediction error exceeds the benign error threshold. All detectors are calibrated to the same false-positive rate, $\alpha=0.02$, making their boundaries comparable. Section~\ref{sub:ml_bdd} describes how each fitted boundary is converted into a scalar stealth constraint, and Section~\ref{sub:comparative_analysis} presents the resulting comparison.

%% file: sections/Problem_Formulations.tex
\section{Formal Modeling of Optimal FDI Attack}
\label{sec:problem_formulation}
To develop our FDI attack analytics framework, we discretize the nonlinear WAMPAC dynamics (Section~\ref{sec: system_dynamics}) using the Backward Euler method for numerical stability. We formulate stealthy FDI as a \rev{MILP} in which the attacker optimizes measurement manipulations to disrupt state estimation and control while evading anomaly detection\rev{; the program is solved with Gurobi}.

\subsection{ADM Constraints for Controllers}
\label{sub: BDD}
To ensure stealth, the ADM restricts \rev{the per-cycle change of each monitored channel to the range it accepts as benign. Because MILP formulations scale poorly for large grids~\cite{jafari23tifs}, the stealth constraints are kept continuous through the slack variables $\epsilon_{\omega}, \epsilon_{v}, \epsilon_{wac}$; binary variables arise only from selecting the accessible controllers (Section~\ref{sub:adv_attr}) and from linearizing the absolute values in the objective~\eqref{eq:attacker-objective}.} System quantities (e.g., power, voltage) are kept within NERC operating limits~\cite{NERC2016}.


\begin{align}
\epsilon^\mathcal{B}_\omega[t], \; \epsilon^\mathcal{B}_v[t], \; \epsilon^\mathcal{B}_{wac}[t] &\geq 0 \label{eq_bdd_1} \\
\omega^b[t] + \epsilon^b_\omega[t] &\geq \bar{\tau}_\omega^b \\
\mathcal{V}^b[t] + \epsilon^b_v[t] &\geq \bar{\tau}_v^b \\
\tau^b_\omega[t] \geq \epsilon^b_\omega[t] &\geq \left|\bar{\omega}^b[t{+}c_\omega] - \bar{\omega}^b[t]\right| \label{eq_frequency_constraints} \\
\tau^b_v[t] \geq \epsilon^b_v[t] &\geq \left|\bar{\mathcal{V}}^b[t{+}c_v] - \bar{\mathcal{V}}^b[t]\right| \\
\tau^b_{wac}[t] \geq \epsilon^b_{wac}[t] &\geq \left|\bar{\mathcal{V}}_{wac}^b[t{+}c_{wac}] - \bar{\mathcal{V}}_{wac}^b[t]\right| \label{eq_wac_constraints} \\
P^b_{m,\min} \leq P^b_m[t] &\leq P^b_{m,\max} \label{eq_power_const} \\
P^b_{g,\min} \leq P^b_g[t] &\leq P^b_{g,\max} \\
V^b_{t,\min} \leq V^b_t[t] &\leq V^b_{t,\max} \label{eq_voltage_const} \\
ACE^i_{\min} \leq ACE^i[t] &\leq ACE^i_{\max} \label{eq_ace_const}
\end{align}

The equations above define the constraints for the ADM framework of our proposed model. $\epsilon^b_{\omega}[t], \epsilon^b_{v}[t], \epsilon^b_{wac}[t] $ ensure non-negativity of auxiliary variables which capture deviations in system parameters such as frequency, voltage, and wide-area control signals. Constraints~\eqref{eq_frequency_constraints}--\eqref{eq_wac_constraints} \rev{cap the change of each compromised channel over its reporting cadence $c_\chi$ by the ADM threshold $\tau_\chi$, so that every injected step resembles a benign variation between consecutive reports; time is counted in time steps (TS).} Operational limits on \rev{mechanical power, electrical power,} and terminal voltage are enforced in~\eqref{eq_power_const}--\eqref{eq_voltage_const}. Tie-line power imbalances are bounded by ACE constraints in ~\eqref{eq_ace_const} to maintain system stability while adhering to AGC's monitoring requirements. \rev{Together, these constraints confine the attack to the region the detector accepts as benign while keeping the grid within its operating limits.}

\subsection{Data-Driven ADM Constraints}
\label{sub:ml_bdd}
The bounds in~\eqref{eq_frequency_constraints}--\eqref{eq_wac_constraints} are
parameterized by the thresholds $\tau_\omega, \tau_v, \tau_{wac}$, which a static BDD fixes once at commissioning time from engineering margins. A modern control center instead calibrates its detector on recorded operation, so the thresholds that the attacker must respect are a function of the benign data. We therefore treat the detector as a parameter in the analysis and derive the stealth bounds from the benign trajectory produced by the same model, thereby keeping the defender and the attacker consistent with one another.

Let $\chi \in \{\omega, v, wac\}$ index the three monitored channels and let
$\Delta_\chi^b[t] = |\bar{\chi}^b[t{+}c_\chi] - \bar{\chi}^b[t]|$ \rev{denote} the
per-controller rate of change over the cadence $c_\chi$ at which that channel is reported, namely one AGC cycle for frequency and one AVR cycle for terminal
voltage and for the wide-area signal. Collecting $\Delta_\chi$ over the benign
run yields the calibration sample $\mathcal{D}_\chi$. A detector $\mathcal{M}_\chi: \mathbb{R} \to \{0,1\}$ is fitted to $\mathcal{D}_\chi$ and tuned so that a common fraction $\alpha$ of the benign sample is flagged, which fixes a single benign false-positive rate across every model considered and makes their boundaries directly comparable. The stealth bound handed to the optimizer is then the detection boundary of $\mathcal{M}_\chi$, that is, the smallest per-cycle excursion the detector still reports as anomalous:
\begin{equation}
\label{eq:ml_threshold}
\small \tau_\chi^{\mathit{ML}} = \max\!\Big( \min\big\{ r \in \mathcal{R} : \mathcal{M}_\chi(r) = 1 \big\}, \, \tau_\chi^{\min}, \, \kappa \max \mathcal{D}_\chi \Big)
\end{equation}
where $\mathcal{R}$ is a probe grid of candidate excursion magnitudes,
$\tau_\chi^{\min}$ is a numerical floor, and the term $\kappa \max \mathcal{D}_\chi$
with $\kappa = 1.05$ guarantees that the benign trajectory itself satisfies the learned bound. Without that last term, a detector calibrated on the mean channel can fall below the excursion of an individual controller, in which case an uncompromised controller would violate its own ADM constraint, and the budget-constrained attack problem would become spuriously infeasible.

Because grid conditions are not stationary, the bounds can also be recalibrated on a sliding basis. Partitioning the horizon $\mathcal{T}$ into windows $W_1, \dots, W_n$ of fixed length \rev{$L_W$ (10 TS)} and applying~\eqref{eq:ml_threshold} to $\mathcal{D}_\chi \cap W_k$ yields a piecewise-constant threshold $\tau_\chi^{\mathit{ML}}[k]$ that widens while the system is riding through a disturbance and tightens again in steady state. This is the adaptive detector the evaluation uses, and~\eqref{eq_frequency_constraints}-\eqref{eq_wac_constraints}
are enforced against $\tau_\chi^{\mathit{ML}}[k]$ for every $t \in W_k$. The static BDD is recovered as the special case in which $\mathcal{M}_\chi$ is the fixed rule and $\tau_\chi^{\mathit{ML}}[k]$ is constant, so both detection paradigms are evaluated by the identical optimization problem and differ only in this parameter.

\subsection{Attack Technique}
Let $\tilde{\omega}_{t}^{b}$ and $\tilde{\mathcal{V}}_{t,t}^{b}$ denote the malicious attack vectors injected into the frequency and voltage measurements utilized by the AGC and AVR, respectively, while $\tilde{\omega}_{wac,t}^{b}$ represents the payload targeting the PMU data streams consumed by the WAMPAC. If $\bar{\omega}_{t}^{b}$, $\bar{\mathcal{V}}_{t,t}^{b}$, and $\bar{\omega}_{wac,t}^{b}$ denote the resulting compromised measurements, the FDI mapping is formally expressed as: 
\begin{align}
\bar{\omega}_t^b &= \omega_t^b + \tilde{\omega}_t^b, \quad &\forall b\in\mathcal{B}, t\in\mathcal{T} \label{eq:frequency-attack} \\
\bar{\mathcal{V}}_{t,t}^b &= \mathcal{V}_{t,t}^b + \tilde{\mathcal{V}}_{t,t}^b, \quad &\forall b\in\mathcal{B}, t\in\mathcal{T} \label{eq:voltage-attack} \\
\bar{\omega}_{wac,t}^b &= \omega_{wac,t}^b + \tilde{\omega}_{wac,t}^b, \quad &\forall b\in\mathcal{B}, t\in\mathcal{T} \label{eq:wac-attack}
\end{align}

\noindent \textbf{\rev{Attack Execution:}}
\rev{The adversary first passively intercepts unencrypted IEEE C37.118 synchrophasor streams through exposed edge devices to learn the benign operating transients. Having identified exploitable PMU nodes and routing weaknesses~\cite{Swain24Access}, it then uses man-in-the-middle techniques (e.g., ARP poisoning and packet crafting) to overwrite the legitimate telemetry in real time with the synthesized vectors, realizing~\eqref{eq:frequency-attack}--\eqref{eq:wac-attack}.}

\subsection{Attacker Goal}
\rev{The attacker's} physical objective is to maximize the deviation of the system's frequency and voltage from their nominal steady-states ($\omega_s, \mathcal{V}_s$) until they breach the threshold limits of the protective relays ($\omega_{\text{max}}, \omega_{\text{min}}, \mathcal{V}_{\text{max}}, \mathcal{V}_{\text{min}}$). Analytically, this is formulated as an optimization problem maximizing the \rev{cumulative absolute deviation} of the grid state:
\begin{equation}
\label{eq:attacker-objective}
\max_{\tilde{\omega}, \tilde{\mathcal{V}}} \sum_{t \in \mathcal{T}} \sum_{b \in \mathcal{B}} \left( \left| \bar{\omega}^b_t - \omega_s \right| + \left| \bar{\mathcal{V}}^b_{t,t} - \mathcal{V}^b_s \right| + \left| \bar{\omega}^b_{wac,t} - \omega_s \right| \right)
\end{equation}

\rev{Maximizing this objective drives the grid state toward, and keeps it in,} the hazardous operating zone long enough to actuate the relays, \rev{which can initiate} localized disconnections or cascading blackouts. \rev{As~\eqref{eq:attacker-objective} is sign-agnostic, the attack may drive the system toward under- or over-frequency.}

\subsection{Adversarial Attributes}
\label{sub:adv_attr}
The attack surface is constrained by the adversary's network accessibility and synthesis capabilities. \textit{Accessibility} refers to the physical or logical reachability of specific measurement nodes, determined by the grid's cybersecurity posture (e.g., firewalls, network segmentation). It is modeled by one boolean vector per channel, $\mathbb{A}^{\chi} \in \{0,1\}^{|\mathcal{B}|}$ with $\chi \in \{agc, avr, wac\}$, whose sum is limited by an accessibility budget. If the attacker has not compromised channel $\chi$ at bus $b$, the corresponding measurement is immutable:
\begin{align}
\textstyle\sum_{b \in \mathcal{B}} \mathbb{A}^{\chi}_b &\leq \beta_{\chi}, \quad \chi \in \{agc, avr, wac\} \label{eq:budget}\\
\mathbb{A}^{agc}_b = 0 &\implies \tilde{\omega}_t^b = 0, \quad \mathbb{A}^{avr}_b = 0 \implies \tilde{\mathcal{V}}_{t,t}^b = 0, \nonumber\\
\mathbb{A}^{wac}_b = 0 &\implies \tilde{\omega}_{wac,t}^b = 0, \quad \forall b \in \mathcal{B}, t \in \mathcal{T}. \label{eq:access}
\end{align}
{\revb On the 118-bus system, jointly optimizing the accessible set and the injections is combinatorial. We therefore fix $\mathbb{A}$ in a preprocessing stage and let the MILP optimize only the injections over that set. Since $\beta_{agc}=10$ already covers every governor-equipped unit, only the wide-area set requires selection. Restricting the attacker's choice of controllers in this way can only lower the attainable objective, so the reported values are lower bounds on the severity achievable under the same budget.}

\subsection{Attack Strategies and ADM Evasion}
Executing the objective in \eqref{eq:attacker-objective} requires a trajectory-aware strategy. A single-shot injection of maximal disruption would immediately violate the ADM \rev{rate-of-change} constraints. \rev{Instead, the attacker solves the constrained optimization over the full horizon, injecting a sequence of increments that each stay within the ADM bound} and mimic legitimate load transients~\cite{CAI20231474, Yuan20TPS}. \rev{The attack thus exploits the margin between the per-cycle tolerance of the ADM and the cumulative excursion needed to reach the relay thresholds.}

\subsection{Attack Assumptions}
The proposed framework considers the following assumptions:
\begin{itemize}
\item \textbf{\textit{Assumption I:}} The \rev{attacker} has adequate knowledge of the AGC, AVR, WAMPAC properties, control algorithm, and ADM constraints. Moreover, the parameters considered in the generators are known to the attacker.
\item \textbf{\textit{Assumption II:}} \rev{Access to a PMU stream implies that the attacker can modify the $V_t$ and $\omega$ measurements of the corresponding bus.}   
\item \textbf{\textit{Assumption III:}} \rev{The attacker has full access to a limited set of PMU and IED measurements (e.g., $V_t$ and $\omega$ of generators and buses), bounded by~\eqref{eq:budget}.} However, control signals sent to and from the AGC and WAMPAC are considered protected against FDI attacks.
\item \textbf{\textit{Assumption IV:}} \rev{The attacker may inject either throughout the horizon or within a bounded window.} 
\end{itemize}

%% file: sections/Evaluation.tex
\section{Evaluation }
\label{sec:evaluation}

\subsection{Hardware-in-the-Loop Setup}
\label{sub:hil_setup}

\begin{figure}[t]
    \centering
    \includegraphics[trim={4cm 4cm 4cm 2.5cm }, clip, width=0.9\columnwidth]{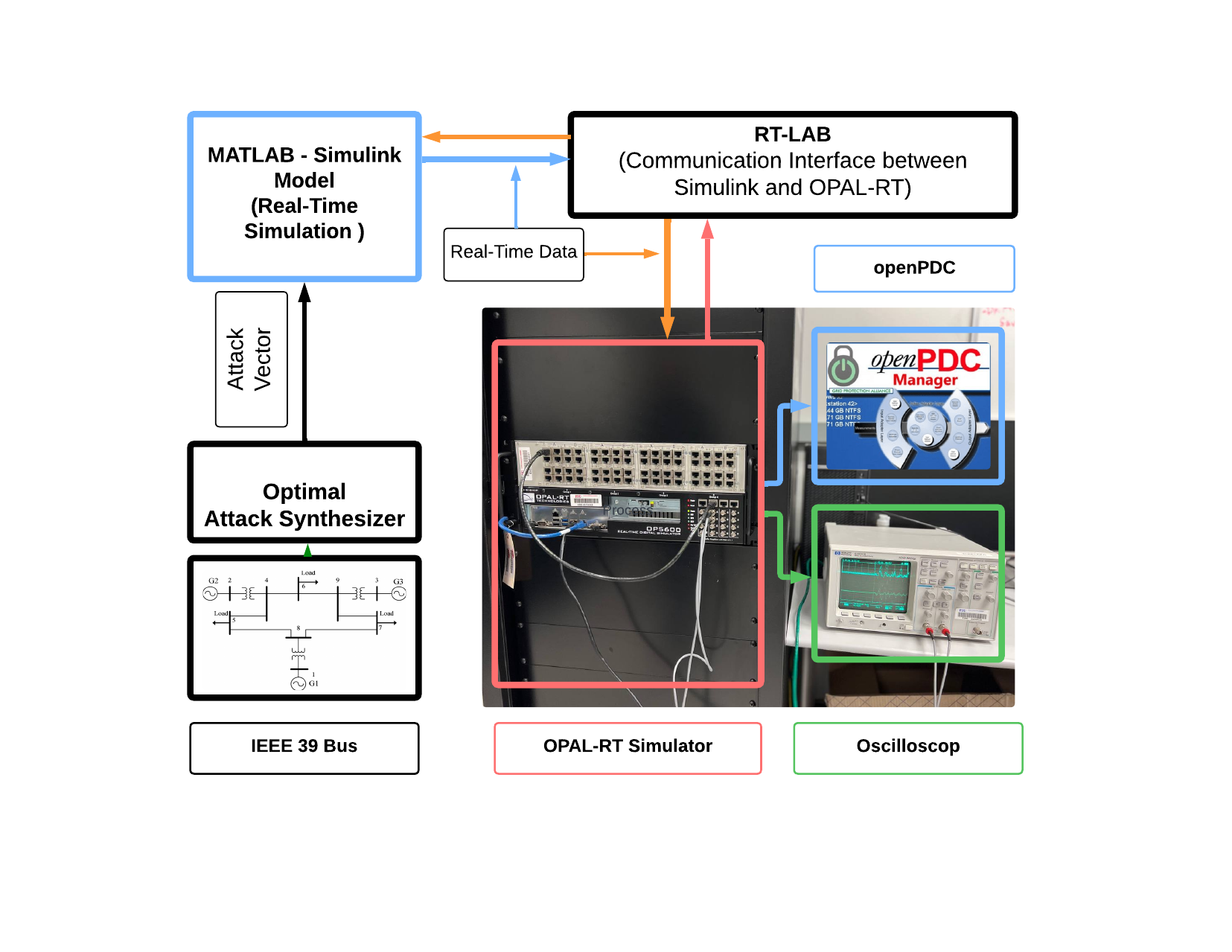}
    \caption{\small Testbed setup of linear optimizer with Simulink and OPAL-RT.}
    \vspace{-6pt}
    \label{plot:testbed}
\end{figure}

Our primary evaluations are conducted on the IEEE 39-bus system, and we then extend the methodology to the larger IEEE 118-bus system to demonstrate scalability. The 39-bus network comprises 10 synchronous generators (total capacity of 10,000~MVA, \rev{load-damping coefficient $D^b = 0.3$ in~\eqref{eq:ace}}) and a constant 6,150~MW load\rev{, partitioned into three control areas}. Generators at buses 30-35 and 39 participate in AGC via governors. All generators utilize AVR and PSS for voltage control except buses 36--38, which maintain constant mechanical input power. Bus~39 serves as the slack bus, and PMUs at all generator terminals stream operational data to the WAMPAC.

To assess the real-time physical impact of optimal FDI attacks, we deployed these models within an OPAL-RT hardware-in-the-loop (HIL) testbed using the ePHASORSIM module (Fig.~\ref{plot:testbed}). By integrating openPDC~\cite{openPDC} via the IEEE C37.118 protocol to manage real-time phasor data, the platform ensures high-fidelity, standards-compliant emulation of cyber-physical grid operations and vulnerabilities. The IEEE 118-bus system extends the evaluation to an industrial scale. It contains 54 synchronous generators across 3 control areas\rev{, of which 10 are governor-equipped and participate in AGC}, with the same WAMPAC PID controller architecture and ADM constraints applied. All the codes and data are publicly available and can be reproduced using \url{https://github.com/ACyD-Lab/Attack-Resiliency-Analytics}.

\begin{table}[t]
\centering
\caption{Simulation Parameters}
\begin{tabular}{|c|c|c|c|}
\hline
\textbf{Parameter} & \textbf{Value} & \textbf{Parameter} & \textbf{Value} \\ \hline
$\mathcal{F}_s$ (Hz) & 60.00 & $\overline{\tau_{\omega}}$ (p.u.) & 1.008 \\ \hline
$\Delta T$ (s) & 1/60 & $\overline{P}_{m,max}^b$ (p.u.) & 1.10 \\ \hline
$S^b$ (MVA) & 1.00 & $ACE^b$ (p.u.) & $\pm$0.05 \\ \hline
$\overline{\tau_{v}}$ (p.u.) & 1.10 & $\overline{\tau_{wac}}$ (p.u.) & 1.008 \\ \hline
$\mathcal{K}^{b}_{P\text{-}wac}$ & 0.01 & $\mathcal{K}^{b}_{P\text{-}avr}$ & 0.75 \\ \hline
$\mathcal{K}^{b}_{I\text{-}wac}$ & 0.01 & $\mathcal{K}^{b}_{I\text{-}avr}$ & 1 \\ \hline
$\mathcal{K}^{b}_{D\text{-}wac}$ & 0.01 & $\mathcal{K}^{b}_{D\text{-}avr}$ & 0.25 \\ \hline
\end{tabular}
\label{tab:simulation_parameter}
\vspace{-6pt}
\end{table}

All experiments use the following canonical configuration: a 600-1200-step optimization horizon (20\,s at $\Delta T = 1/60$\,s), a load disturbance of 800~MW (40~MW per load bus, 13\% of total load) applied \rev{for about 3.3~s} from $\text{TS}=125$ to $\text{TS}=325$, and attack injection beginning at $\text{TS}=60$ with 60-step intervals. Deviations from this setup are stated explicitly in each subsection. Generators' parameters are taken from \cite{Jafari23TSG, Mali2023} and simulation settings provided in Table~\ref{tab:simulation_parameter}. \rev{We evaluate attacks on AGC, AVR, and WAMPAC while varying the ADM threshold, attack duration, and number of compromised generators, and address the following research questions (RQ):}

\begin{itemize}
    \item \textbf{RQ1}. \textit{How does wide-area control affect grid stability and oscillation damping under load disturbances?}
    (Section~\ref{sub:load_change})

    \item \textbf{RQ2}. \textit{How do attack duration and the targeted control layer affect the physical impact of FDI attacks?}
    (Section~\ref{FDI_AGC_AVR})

    \item \textbf{RQ3}. \textit{How does adversarial access to AGC, AVR, and wide-area control channels affect attack severity and time to objective?}
    (Section~\ref{sub:impact_accessibility})

    \item \textbf{RQ4}. \textit{To what extent do learned ADM boundaries reduce stealthy attack severity relative to static BDD constraints, and how does boundary calibration affect residual risk?}
    (Sections~\ref{sub:comparative_analysis} and~\ref{sub:detector_sensitivity})
\end{itemize}
\subsection{Baseline Validation}
\label{sub:load_change}

\begin{figure}[!htbp]
    \centering
    \vspace{-8pt}
        \subfigure[]{
            \label{fig:fq_without_wampac}
            \includegraphics[width=0.46\columnwidth]{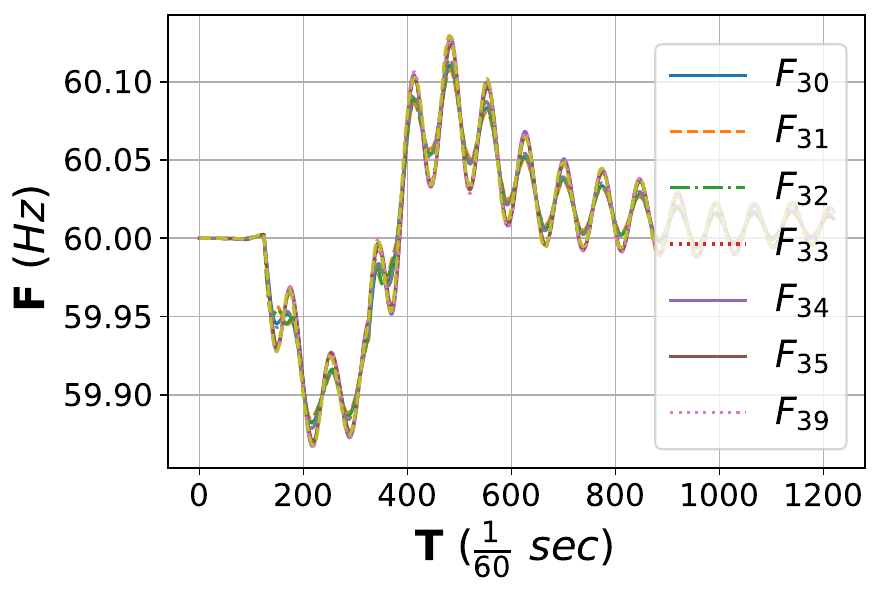}
        }
        \subfigure[]{
            \label{fig:volt_without_wampac}
            \includegraphics[width=0.46\columnwidth]{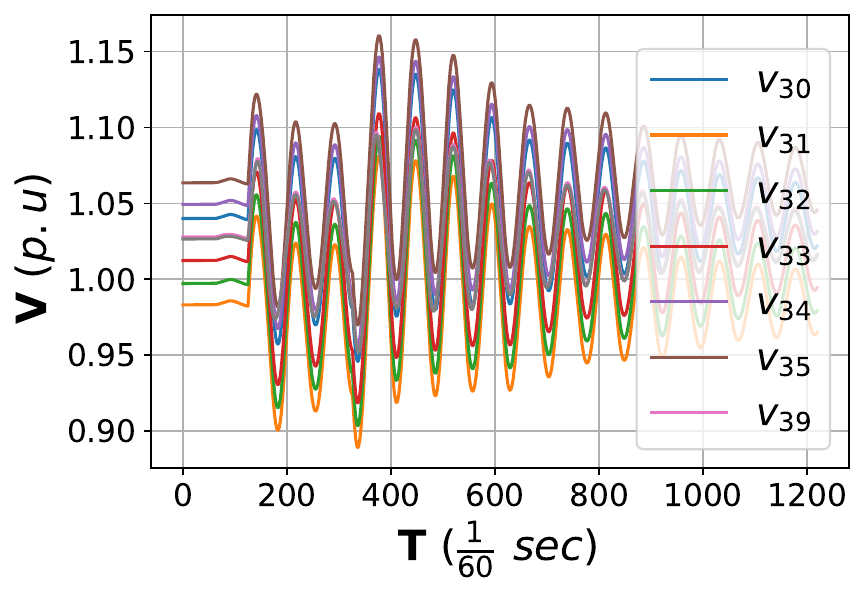}
        }
        \vspace{-8pt}
    \caption{\small Generator response \textit{without} WAMPAC during load disturbance: (a) frequency, (b) voltage.}
    \vspace{-8pt}
    \label{plot:without_wampac_response}
\end{figure}

\begin{figure}[t]
    \centering
        \subfigure[]{
            \label{fig:fq_with_wampac}
            \includegraphics[width=0.46\columnwidth]{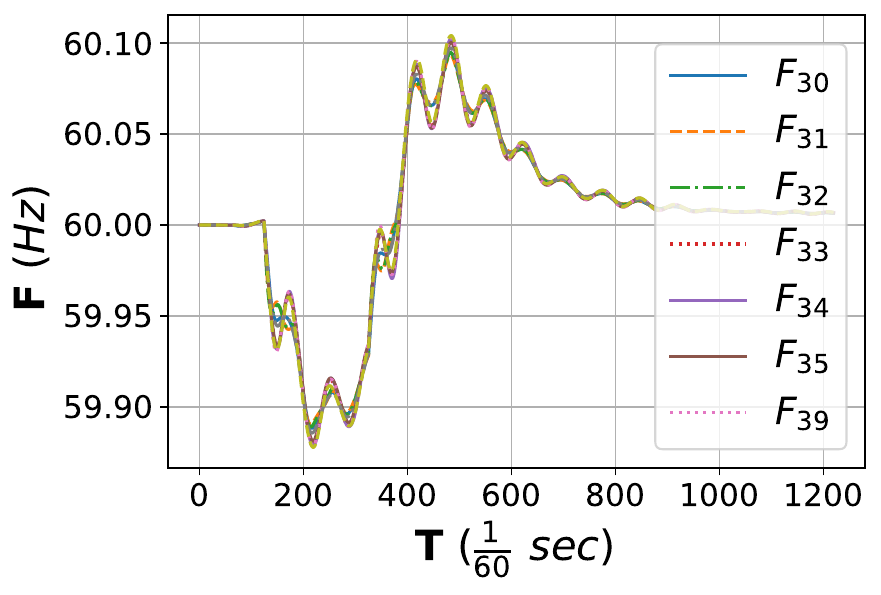}
        }
        \subfigure[]{
            \label{fig:volt_with_wampac}
            \includegraphics[width=0.46\columnwidth]{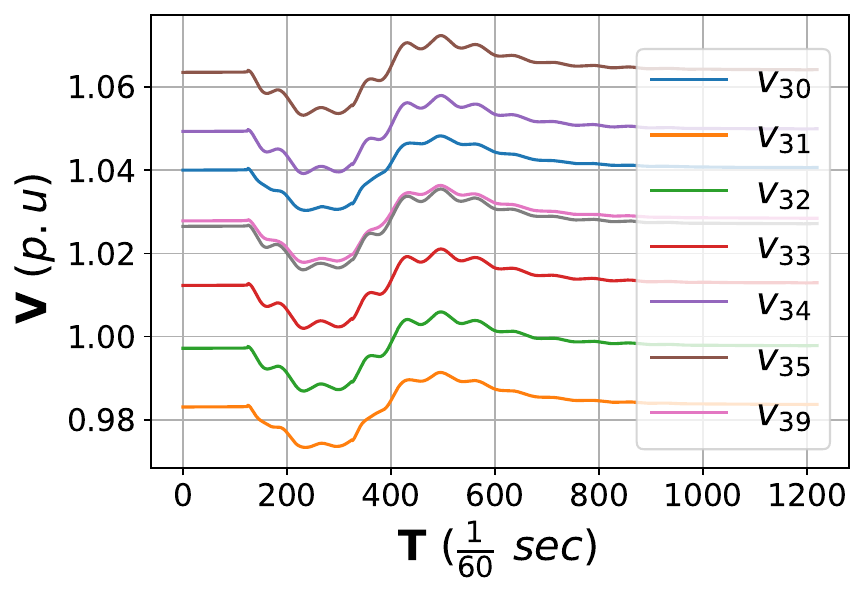}
        }
        \vspace{-6pt}
    \caption{\small Generator response \textit{with} WAMPAC during load disturbance: (a) frequency, (b) voltage.}
    \vspace{-8pt}
    \label{plot:with_wampac_response}
\end{figure}

\begin{figure}[t]
    \centering
        \subfigure[]{
            \label{fig:fq_wampac_gurobi}
            \includegraphics[width=0.46\columnwidth]{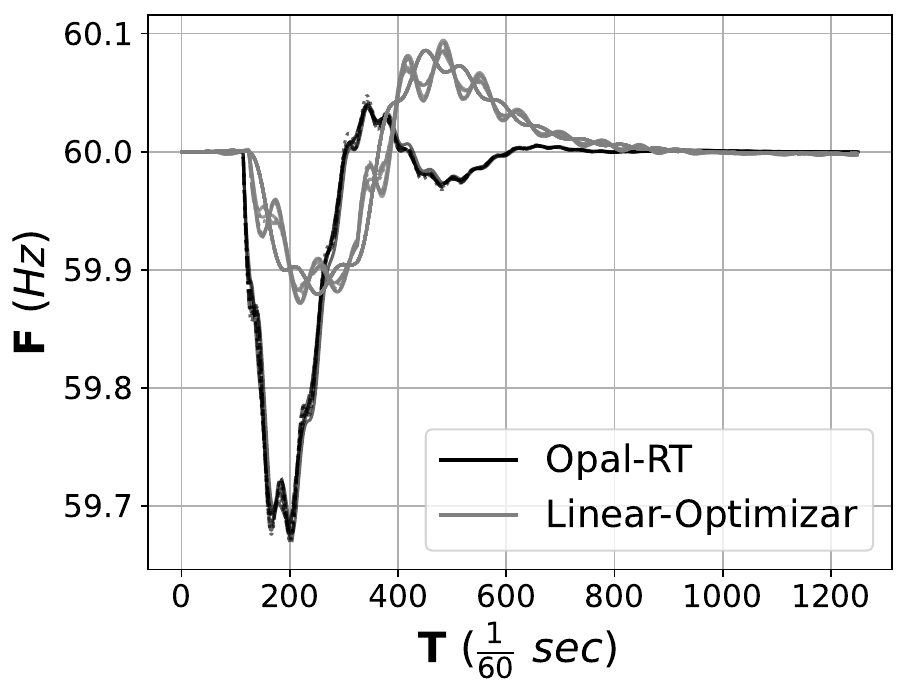}
        }
        \subfigure[]{
            \label{fig:fq_without_wampac_gurobi}
            \includegraphics[width=0.46\columnwidth]{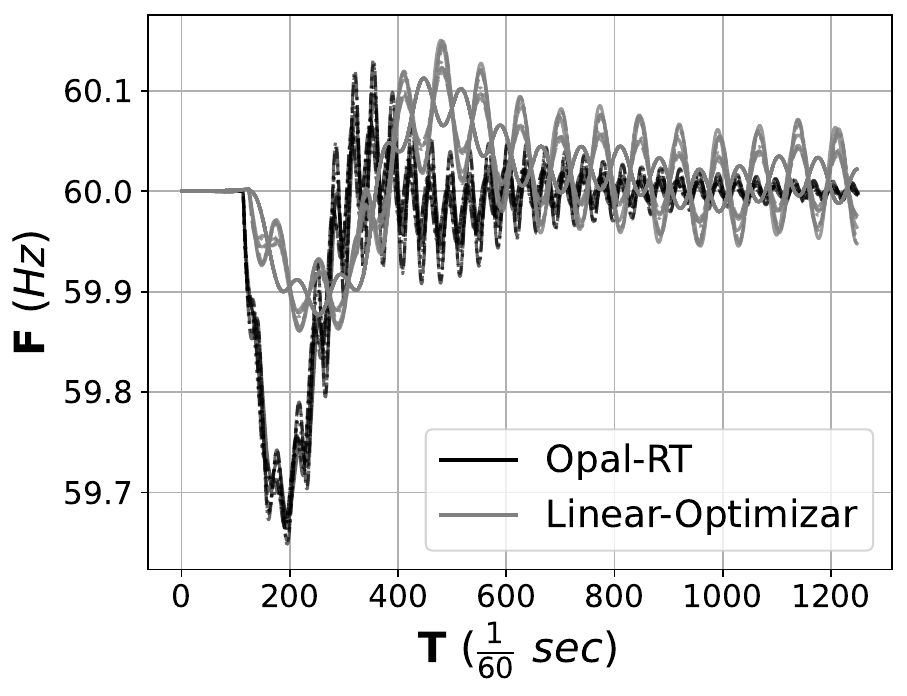}
        }
        \vspace{-6pt}
    \caption{\small Frequency response of linear optimizer vs.\ OPAL-RT model: (a) with WAMPAC, (b) without WAMPAC.}
    \vspace{-8pt}
    \label{plot:wampac_gurobi}
\end{figure}

To verify the fidelity of the power system dynamics, all attack-related constraints were temporarily removed, allowing the system to exhibit its fundamental behavior as modeled in~\eqref{eq:wac_speed_dev}--\eqref{eq:swing}. \rev{The canonical 800~MW disturbance was applied with a time step of 1/60~s ($\approx$0.0167~s).}

Fig.~\ref{plot:without_wampac_response} shows the system response without WAMPAC, while Fig.~\ref{plot:with_wampac_response} shows the response with WAMPAC integrated. The comparative analysis demonstrates that the WAMPAC model reduced the peak-to-peak amplitude of frequency oscillations by approximately 10\% and shortened the settling time through the tie lines. Similar responses were observed on the OPAL-RT testbed (Fig.~\ref{plot:wampac_gurobi}), affirming model fidelity despite minor discrepancies arising from the difference between the linear power-flow method used in the MILP formulation and the AC power-flow method used by ePHASORSIM.


    

\subsection{Impact of Continuous and Timebound FDI Attack}
\label{FDI_AGC_AVR}

\begin{figure}[t]
    \centering
        \subfigure[]{
            \label{fq_with_wampac_1}
            \includegraphics[width=0.46\columnwidth]{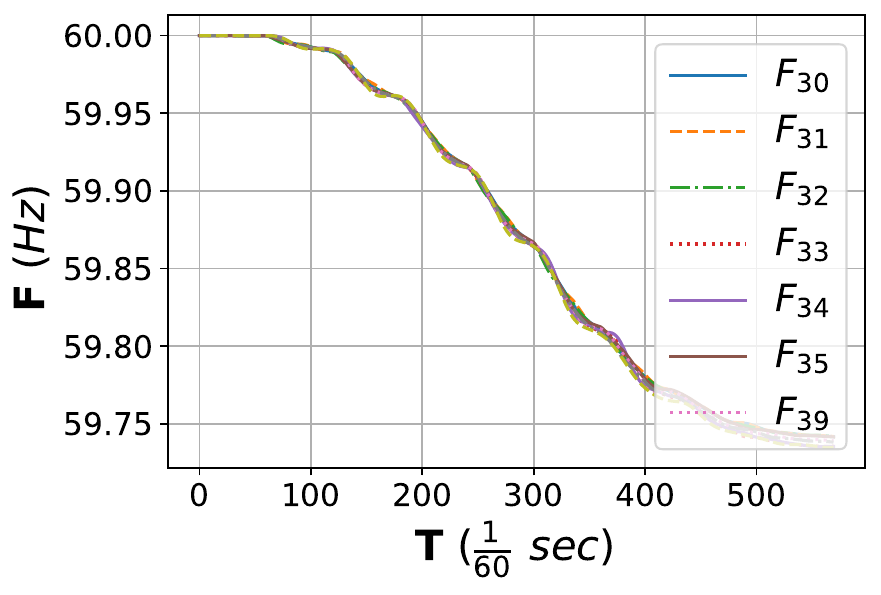}
        }
        \subfigure[]{
            \label{voltage_without_wampac_1}
            \includegraphics[width=0.46\columnwidth]{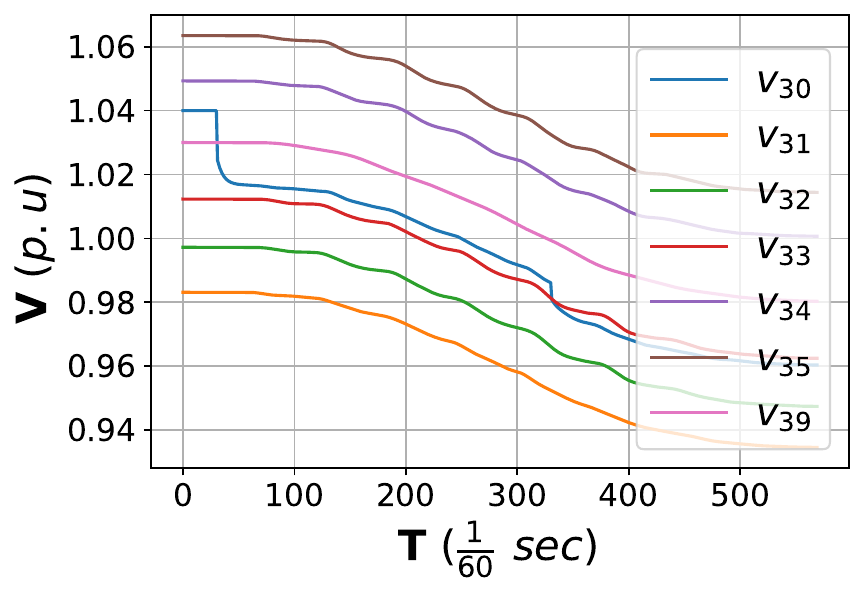}
        }
        \vspace{-6pt}
    \caption{\small Demonstrating generator's response (a) frequency and (b) voltage with the presence of WAMPAC due to continuous attack on AGC.}
    \label{plot:with_wampac_agc_1}
    
\end{figure}
We investigated the impact of continuous and time-bound FDI attacks on local controllers within a WAMPAC framework and presented the findings through \rev{two} test cases outlined below.

\noindent\textbf{Case 1: } We conducted an FDI attack on AGC's frequency measurements received from IEDs and RTUs. The attack involved adding false data to the measurements of each AGC across three different areas at regular intervals corresponding to every $60^{th}$ TS, starting with the experiment's $100^{th}$ TS. The effect of continuous FDI attacks on AGC frequency measurement in all three sites is depicted in Fig.~\ref{plot:with_wampac_agc_1}. The results show that attacks on AGC significantly affect frequency stability, generating fluctuations of up to 0.25 Hz within the $8^{th}$ AGC cycle. 

\begin{figure}[t]
    \centering
        \subfigure[]{
            \label{fq_with_wampac_2}
            \includegraphics[width=0.46\columnwidth]{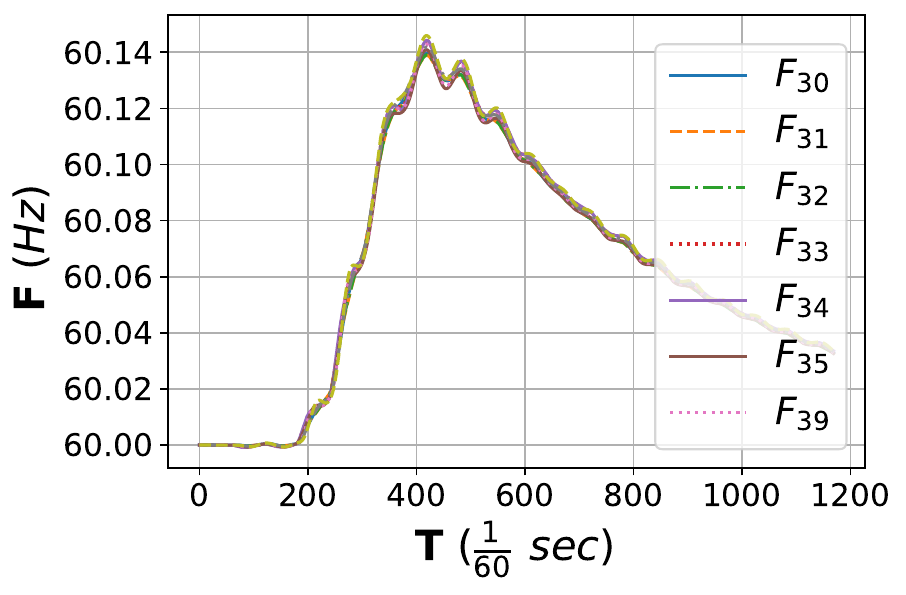}
        }
        \subfigure[]{
            \label{voltage_without_wampac_}
            \includegraphics[width=0.46\columnwidth]{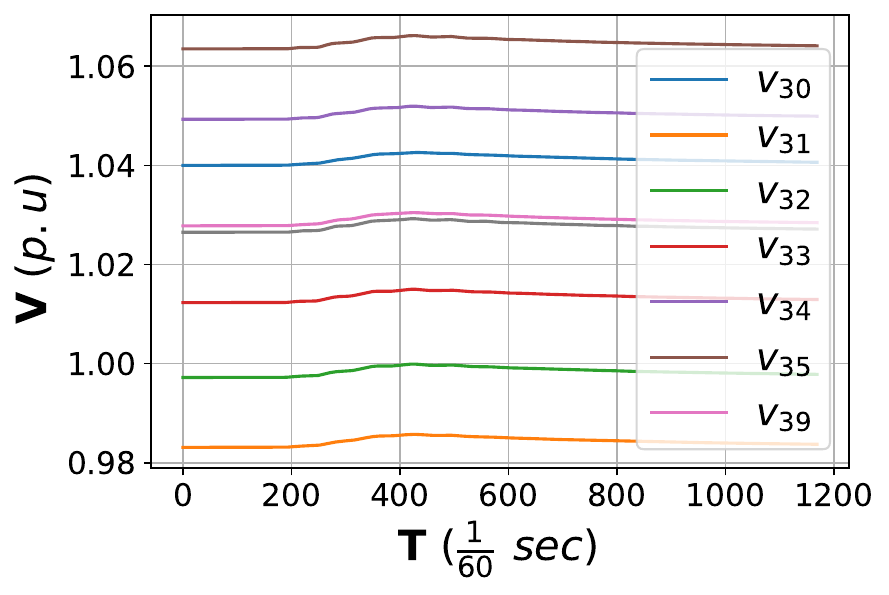}
        }
        \vspace{-6pt}
    \caption{ \small Demonstrating generator's response of (a) frequency and (b) voltage with the presence of WAMPAC due to time-bound attack on AGC.}
    \label{plot:with_wampac_agc_2}
    \vspace{-6pt}
\end{figure}


\noindent\textbf{Case 2: } In this case, malicious data is inserted into AGC frequency measurements at three areas, with a maximum of four AGC cycles. During the simulation, the attacker injected malicious data between the $200^{th}$ and $420^{th}$ TS with 60-step intervals. Fig.~\ref{plot:with_wampac_agc_2} shows the impact of these time-bound FDI attacks on AGC frequency measurements. The results indicate that these attacks cause less severe frequency instability than in \textbf{Case 1}, with variations up to 0.15 Hz. After the attack stops, the system returns to stability. Similar deviation patterns are observed in voltage levels, rotor angles, and tie-line power flows.

\subsection{Analysis of Optimal FDI Attack on WAMPAC}
\label{sub: FDI_WAMPAC}
\begin{figure}[t]
    \centering
        \subfigure[]{
            \label{fq_with_wampac_4}
            \includegraphics[width=0.46\columnwidth]{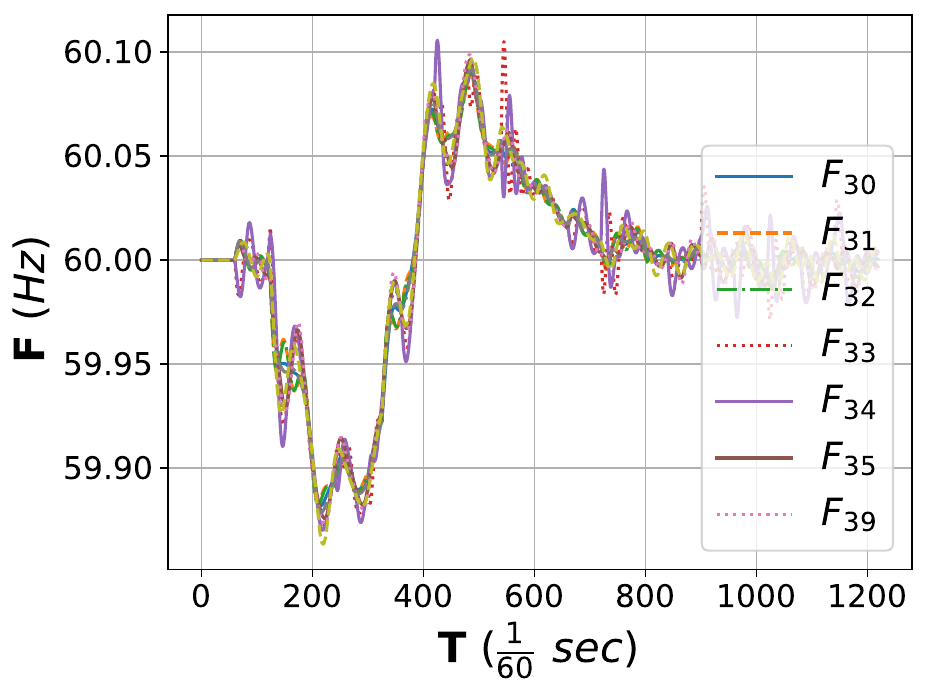}
        }
        \subfigure[]{
            \label{voltage_without_wampac_4}
            \includegraphics[width=0.46\columnwidth]{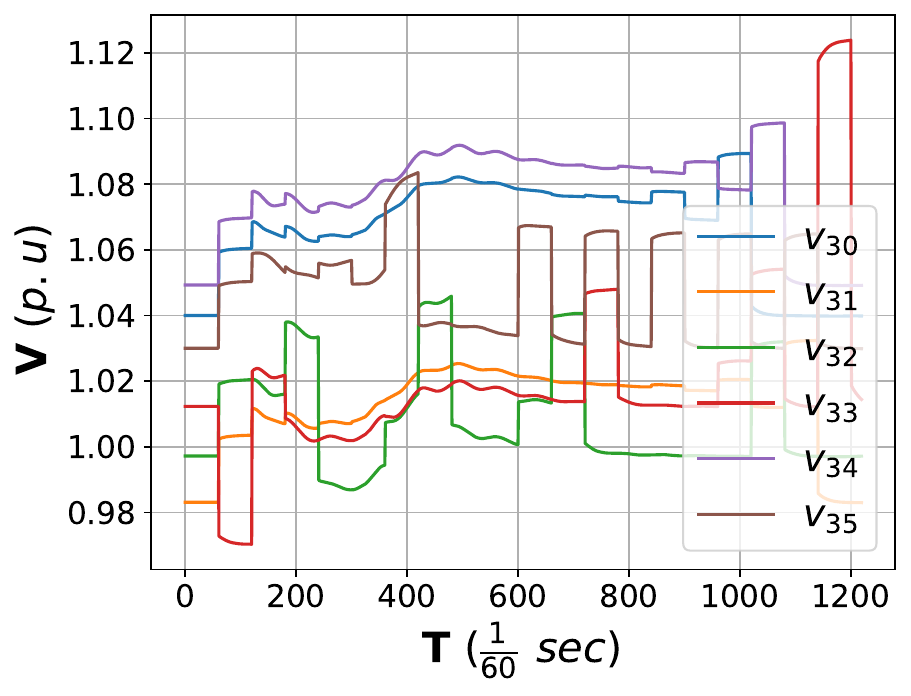}
        }
       \vspace{-6pt}
    \caption{ \small Demonstrating generator's response of (a) frequency and (b) voltage due to attack on WAMPAC during load change}
    \label{plot:with_wampac_agc_4}
    \vspace{-6pt}
\end{figure}

    

In an interconnected power system, WAMPAC is crucial for grid stability, using real-time PMU data to coordinate control actions and dampen oscillations that would otherwise be beyond the reach of local controllers. An attack on WAMPAC, particularly targeting frequency measurements from PMU data, disrupts wide-area coordination, causing greater system frequency and voltage deviations than attacks on local controllers do. \rev{The attack and the load disturbance follow the canonical configuration.} 
Fig.~\ref{plot:with_wampac_agc_4} shows the impact of ongoing FDI attacks on WAMPAC. An attack on WAMPAC disrupts the regular damping control mechanism and introduces oscillations, which causes the system to take 30\% longer to reach a steady state compared to the case in Fig.~\ref{plot:with_wampac_response}.

\subsection{Evaluation of Impact Based on Accessibility}
\label{sub:impact_accessibility}

\begin{figure}[t]
    \centering
        \subfigure[]{
            \label{benign_frequency}
            \includegraphics[width=0.43\columnwidth]{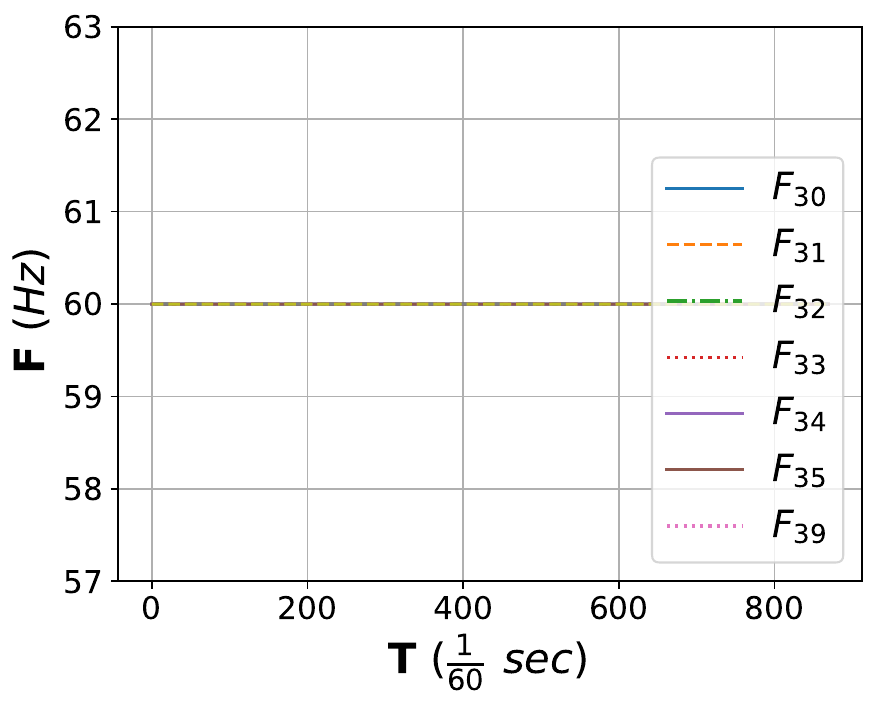}
        }
        \subfigure[]{
            \label{frequency_agc_attck}
            \includegraphics[width=0.46\columnwidth]{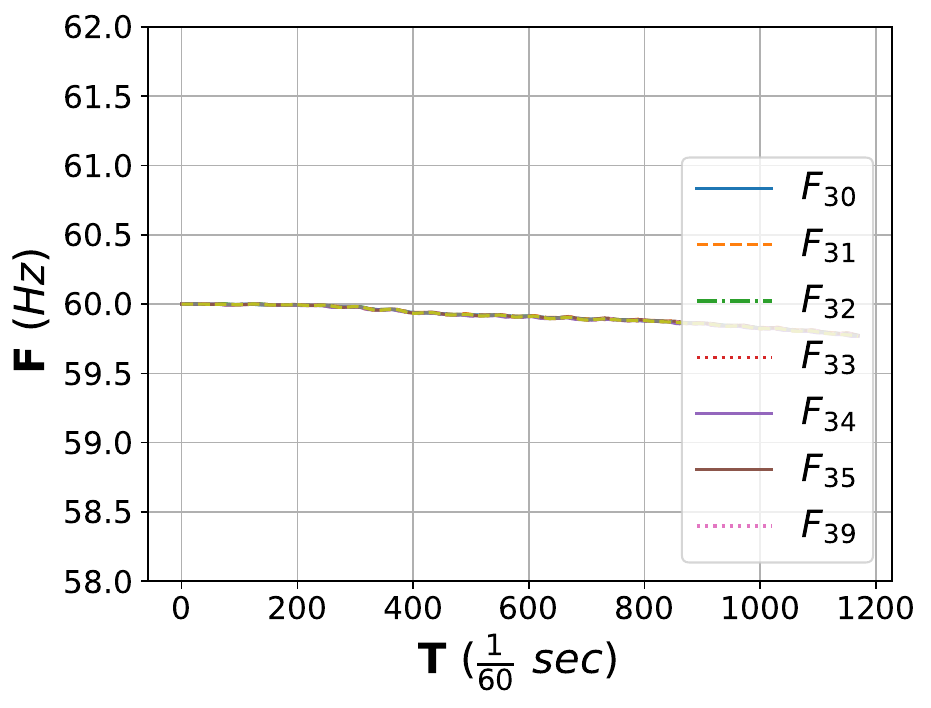}
        } 
        \vspace{-6pt}
    \caption{\small Demonstrating generator's frequency on AGC in (a) benign and (b) attack cases.}
    \label{plot:agc_only}
    \vspace{-6pt}
\end{figure}

\begin{figure}[htbp]
    \centering
        \subfigure[]{
            \label{frequency_agc_avr_attck}
            \includegraphics[width=0.46\columnwidth]{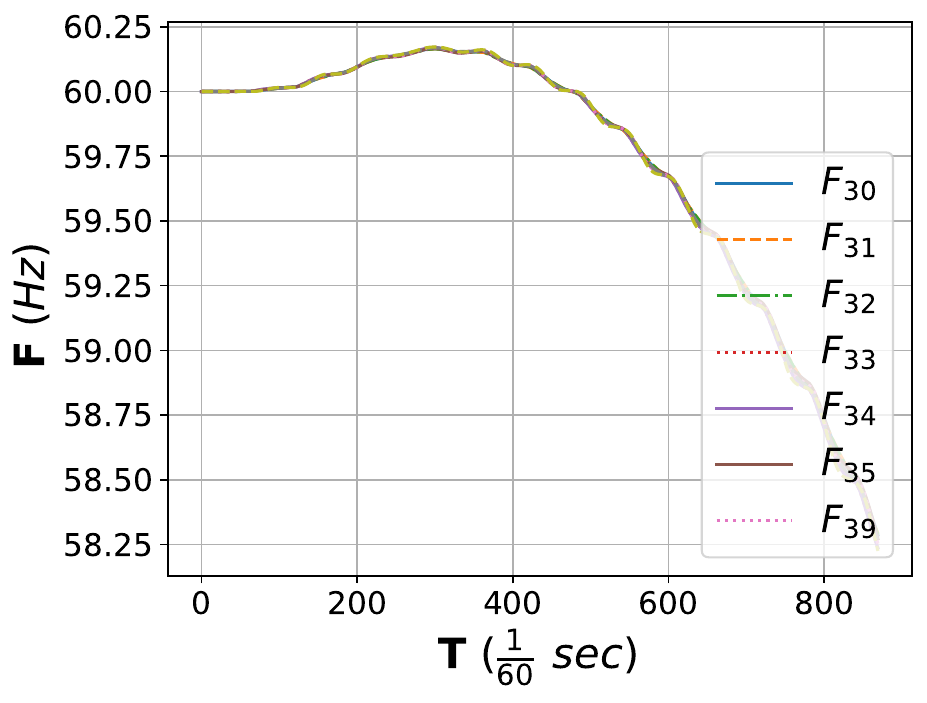}
        }
        \subfigure[]{
            \label{frequency_all_attack}
            \includegraphics[width=0.46\columnwidth]{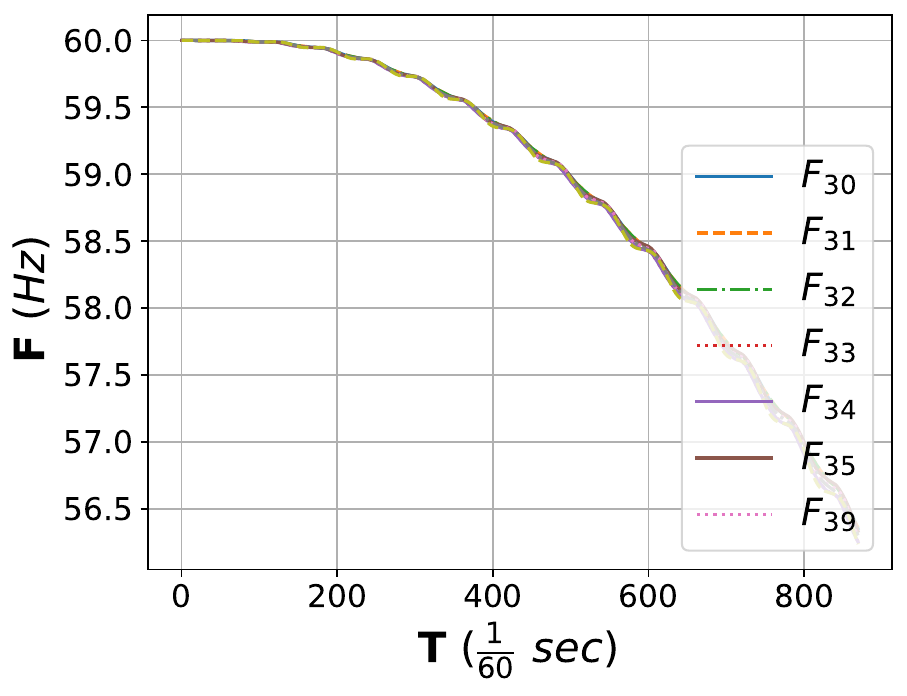}
        }
        \vspace{-6pt}
    \caption{\small Demonstrating generator's frequency for (a) combined attack on AGC and AVR and (b) attack on all controllers including WAMPAC }
    \label{plot:with_wampac_agc_avr}
    \vspace{-6pt}
\end{figure}

\begin{figure*}[ht]
    \centering
        \subfigure[]{
            \label{fig:39_cd_freq}
            \includegraphics[width=0.30\textwidth]{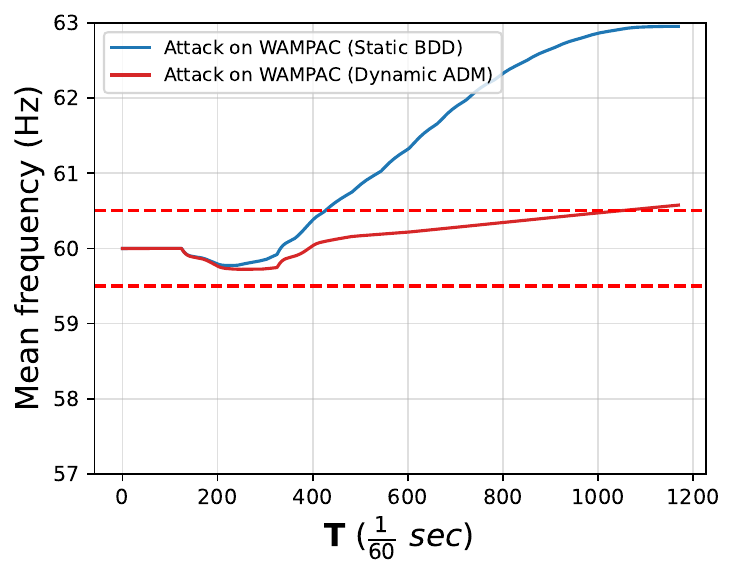}
        }
        \subfigure[]{
            \label{fig:39_cd_tie}
            \includegraphics[width=0.30\textwidth]{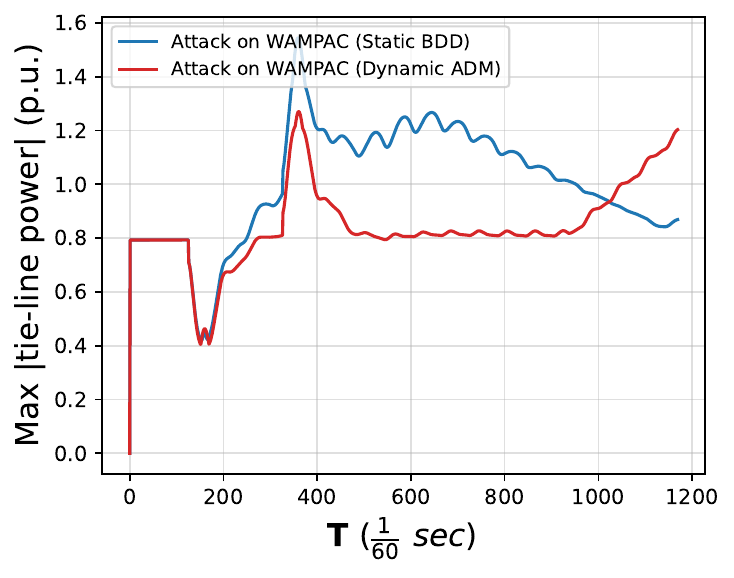}
        }
        \subfigure[]{
            \label{fig:39_cd_volt}
            \includegraphics[width=0.30\textwidth]{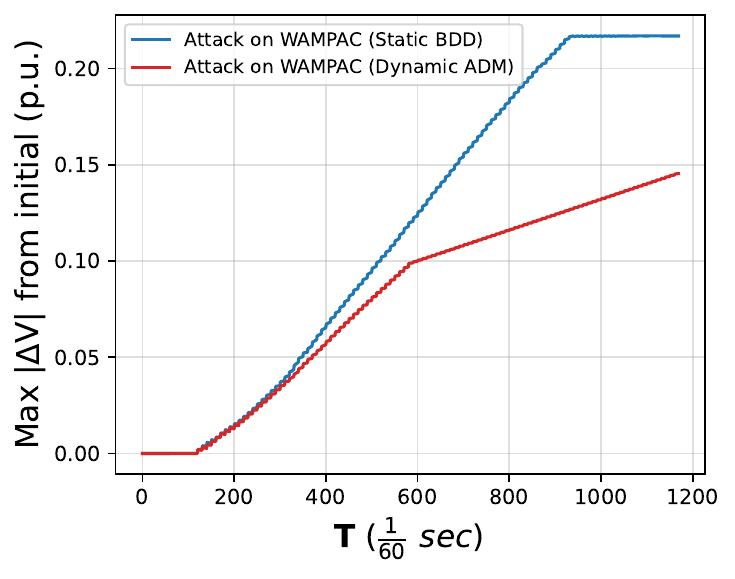}
        }
        \vspace{-6pt}
    \caption{\small IEEE 39-bus system under the static BDD (Scenario~C) and the learned \rev{ADM} (Scenario~D\rev{, labeled "Dynamic ADM"}): (a) generator frequency, (b) tie-line power, (c) terminal voltage deviation.}
    \label{plot:39bus_comparison}
    \vspace{-6pt}
\end{figure*}

\begin{figure*}[ht]
    \centering
        \subfigure[]{
            \label{fig:118_cd_freq}
            \includegraphics[width=0.30\textwidth]{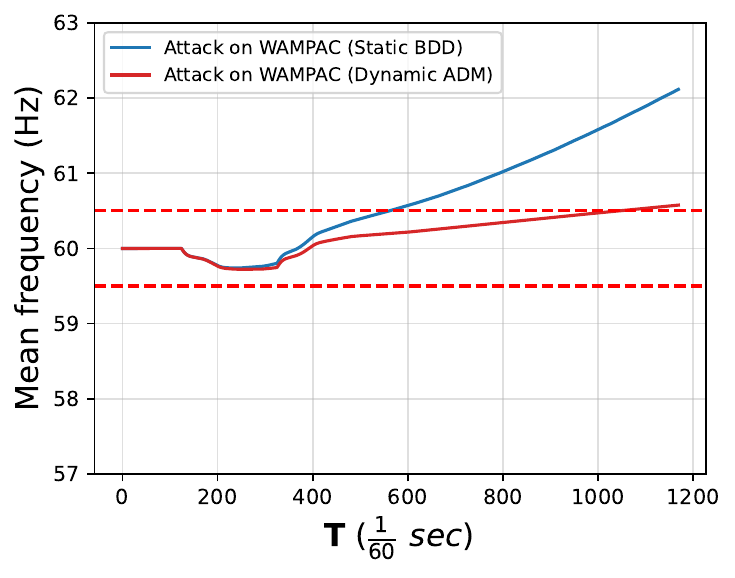}
        }
        \subfigure[]{
            \label{fig:118_cd_tie}
            \includegraphics[width=0.30\textwidth]{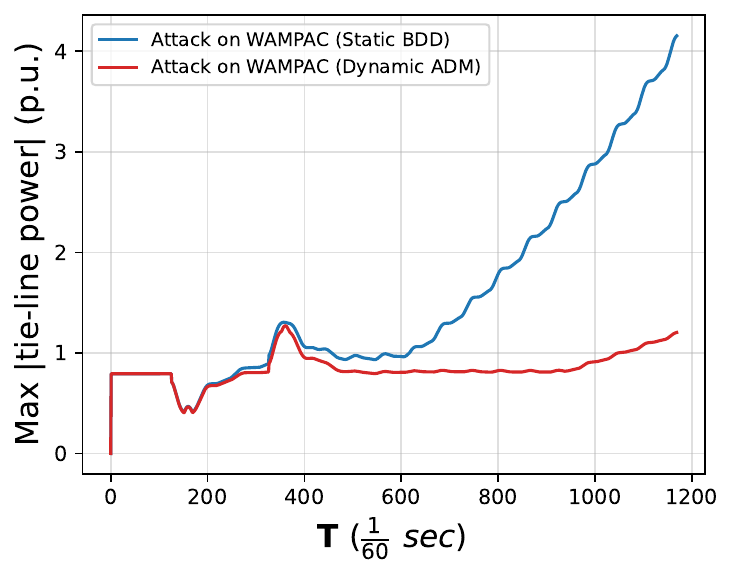}
        }
        \subfigure[]{
            \label{fig:118_cd_volt}
            \includegraphics[width=0.30\textwidth]{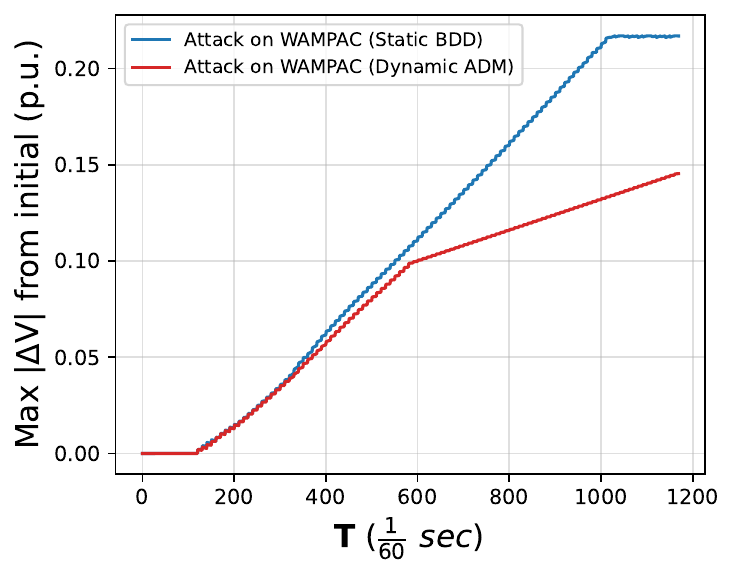}
        }
        \vspace{-6pt}
    \caption{\small IEEE 118-bus system under the static BDD (Scenario~C) and the learned \rev{ADM} (Scenario~D\rev{, labeled ``Dynamic ADM''}): (a) generator frequency, (b) tie-line power, (c) terminal voltage deviation.}
    \label{plot:118bus_comparison}
    \vspace{-6pt}
\end{figure*}

In this scenario, we have demonstrated how an attack on WAMPAC can achieve a specific attack goal. We assume an attacker targets the AGC while the AVR and WAMPAC remain unaffected; it takes \rev{more than 800 TS} to achieve the desired frequency deviation of more than 0.5~Hz, as shown in Fig.~\ref{frequency_agc_attck}. \rev{Because frequency deviations couple into the excitation loop, an uncompromised AVR resists the AGC-only attack and slows it.} When both the AGC and AVR are attacked while WAMPAC remains intact, the attacker reaches the desired deviation in \rev{about 500 TS} (Fig.~\ref{frequency_agc_avr_attck}). When the attacker targets WAMPAC in addition to the AGC and AVR, the system reaches the desired deviation even faster, within \rev{380 TS}.
Fig.~\ref{frequency_all_attack} \rev{shows that a compromised WAMPAC propagates the attack across areas through inter-area coupling.} Attacking WAMPAC has a greater impact on system stability due to its wide-area control role, underscoring its importance in preserving grid stability.

\subsection{Comparative Analysis: Static versus Learned ADM Boundaries}
\label{sub:comparative_analysis}
This subsection evaluates the benefit of replacing static BDD rules with boundaries learned from benign operation. All runs use the same horizon, load step, accessibility budget, and optimization problem; only the ADM constraints change. Scenarios A and B are benign responses with the wide-area loop disabled and enabled, respectively. Scenarios~C and~D apply the optimal attack under the static \rev{BDD and learned ADM} constraints, with Scenario~D calibrated on the corresponding Scenario-B data. Both attacks use the same compromised controllers, isolating the effect of the stealth constraint. The 39-bus budget \rev{($\beta_{agc}/\beta_{wac}=3/7$)} includes three governors and seven \rev{wide-area} generators; the 118-bus budget \rev{($10/9$)} includes all ten governors and nine \rev{wide-area} generators selected \rev{as described in Section~\ref{sub:adv_attr}}.

\begin{table}[htbp]
\centering
\caption{\small Four-scenario comparison on the IEEE 39-bus and 118-bus systems.
}
\label{tab:wampac_scenarios}
\footnotesize
\setlength{\tabcolsep}{3.5pt}
\setlength{\tabcolsep}{2.2pt}\begin{tabular}{lllrrr}
\toprule
\textbf{System} & \textbf{Sc.} & \textbf{Configuration} &
$\boldsymbol{\mathcal{J}}$ & \rev{$\boldsymbol{t_{60.5}}$} &
$\boldsymbol{t_{\mathrm{sol}}}$ \\
\midrule
\multirow{4}{*}{\textbf{39-bus}}
 & A & No WAMPAC, benign            & 241.67  & \rev{--} & 12.0  \\
 & B & WAMPAC, benign               & 242.28  & \rev{--} & 18.0  \\
 & C & \rev{WAMPAC, attack, static BDD} & 787.69  & \rev{400} & 49.7  \\
 & D & \rev{WAMPAC, attack, learned ADM} & 508.72  & \rev{880} & 89.1  \\
\midrule
\multirow{4}{*}{\textbf{118-bus}}
 & A & No WAMPAC, benign            & 289.87  & \rev{--} & 70.9  \\
 & B & WAMPAC, benign               & 290.38  & \rev{--} & 116.2 \\
 & C & \rev{WAMPAC, attack, static BDD} & 2352.34 & \rev{580} & 509.6 \\
 & D & \rev{WAMPAC, attack, learned ADM} & 1058.93 & \rev{1000--1020} & 401.6 \\
\bottomrule
\end{tabular}

\vspace{2pt}
\begin{minipage}{0.98\columnwidth}\footnotesize
Static BDD thresholds are $(\tau_\omega, \tau_v, \tau_{wac}) = (0.007, 0.007,
0.70)$~p.u. \rev{A dash means the setting is not reached within the 1200-TS horizon.} $\mathcal{J}$ is the attained MILP objective, \rev{$t_{60.5}$ the first time step (TS) at which the mean generator frequency reaches the 60.5~Hz over-frequency relay setting,} and $t_{\mathrm{sol}}$ the solver
wall-clock time \rev{(s)}. \rev{For the benign Scenarios~A and~B, $\mathcal{J}$ is~\eqref{eq:attacker-objective} evaluated on the benign trajectory. Solve times are the mean over the repeated solves that share an identical configuration across the detector runs of Table~\ref{tab:static_vs_ml_bdd}.}
\end{minipage}
\end{table}

Table~\ref{tab:wampac_scenarios} shows that the wide-area loop has little effect on the benign aggregate frequency metric: \rev{no benign run reaches the 60.5~Hz setting within the horizon, and enabling the loop changes $\mathcal{J}$ by less than 0.3\%}. Its damping benefit instead appears in the voltage, rotor-angle, and settling responses reported in Section~\ref{sub:load_change}. \rev{With wide-area access, however, the optimal attack attains $3.3\times$ (39-bus) and $8.1\times$ (118-bus) the benign objective (Scenario~C versus~B)}, because a single wide-area signal reaches the excitation systems of multiple generators. Thus, its value to an attacker is substantially greater than its apparent contribution to the benign frequency metric.

Under the static BDD, the optimal stealthy attack \rev{drives the mean generator frequency to the 60.5~Hz over-frequency setting at 400~TS on the 39-bus system and at 580~TS on the 118-bus system. Learned bounds reduce the objective by 35.4\% and 55.0\%, respectively, and postpone the crossing to 880~TS and 1000--1020~TS} by limiting the rate at which deviation accumulates (Figs.~\ref{plot:39bus_comparison} and~\ref{plot:118bus_comparison}). {\revb The tie-line flow shows the same effect: on the 118-bus system, the static-BDD attack drives its peak above 4~p.u., whereas the learned ADM holds it near 1.2~p.u.\ (Fig.~\ref{fig:118_cd_tie}); on the 39-bus system, the early peak drops from about 1.5 to 1.3~p.u.\ (Fig.~\ref{fig:39_cd_tie}).} {\revb Nevertheless, the learned bound delays the trip rather than preventing it: every Scenario-D attack still reaches 60.5~Hz within the horizon with an objective exceeding twice the benign baseline, and the terminal-voltage excursion also exceeds the $\pm 10\%$ band of Section~\ref{sec:technical} (Figs.~\ref{fig:39_cd_volt} and~\ref{fig:118_cd_volt}). The learned detector therefore buys the operator about 8~s (39-bus) and 7~s (118-bus) of response time rather than eliminating the attack surface.}

\subsection{Sensitivity to the Learned Detection Boundary}
\label{sub:detector_sensitivity}

\begin{table*}[t]
\centering
\caption{\small Per-detector comparison of the static-BDD attack (Scenario~C) and the
learned-\rev{ADM} attack (Scenario~D) on the IEEE 39-bus and 118-bus systems. }
\vspace{-5pt}
\label{tab:static_vs_ml_bdd}
\footnotesize
\setlength{\tabcolsep}{3pt}
\begin{tabular}{clcrrrrrrrrr}
\toprule
\textbf{Test} & \textbf{Learned} & \textbf{Budget}
& \multicolumn{2}{c}{\textbf{Learned bound} ($\times 10^{-3}$)}
& \multicolumn{3}{c}{\textbf{Attack objective} $\boldsymbol{\mathcal{J}}$}
& \multicolumn{2}{c}{$\boldsymbol{f_{\mathrm{60.5}}}$ \textbf{(TS)}}
& \multicolumn{2}{c}{\textbf{Solve time (s)}} \\
\cmidrule(lr){4-5}\cmidrule(lr){6-8}\cmidrule(lr){9-10}\cmidrule(lr){11-12}
\textbf{system} & \textbf{detector} & $\beta_{\mathit{agc}}/\beta_{\mathit{wac}}$
& $\tau_\omega^{\mathit{ML}}$ & $\tau_v^{\mathit{ML}}$
& \textbf{C} & \textbf{D} & $\boldsymbol{\Delta}$ \textbf{(\%)}
& \textbf{C} & \textbf{D}
& \textbf{C} & \textbf{D} \\
\midrule
\multirow{4}{*}{IEEE 39-bus}
& DBSCAN                  & 3/7   & 3.81  & 2.20 & 787.69  & 508.72  & $-35.4$ & 400 & \textbf{880} & 42.6  & 72.6  \\
& Isolation forest        & 3/7   & 3.81  & 2.20 & 787.69  & 508.72  & $-35.4$ & 400 & \textbf{880} & 58.6  & 106.7 \\
& One-class SVM           & 3/7   & 3.81  & 2.20 & 787.69  & 508.72  & $-35.4$ & 400 & \textbf{880} & 49.5  & 87.9  \\
& LSTM$^{\ddagger}$       & 3/7   & 10.29 & 4.63 & 787.69  & 665.44  & $-15.5$ & 400 & \textbf{1000}          & 48.1  & 40.1  \\
\midrule
\multirow{4}{*}{IEEE 118-bus}
& DBSCAN                  & 11/19  & 2.65  & 0.96 & 2352.34 & 1058.93 & $-55.0$ & 580 & \textbf{1020} & 514.2 & 421.4 \\
& One-class SVM           & 11/19  & 2.65  & 0.96 & 2352.34 & 1058.93 & $-55.0$ & 580 & \textbf{1000} & 501.3 & 381.7 \\
& LSTM$^{\ddagger}$       & 11/19  & 3.78  & 1.18 & 2352.34 & 1140.90 & $-51.5$ & 580 & \textbf{890} & 513.2 & 403.8 \\
& Isolation forest$^{\S}$ & 11/19 & 2.76  & 1.54 & 1700.07 & 1112.96 & $-34.5$ & 580 & \textbf{900}          & 200.9 & 243.1 \\
\bottomrule
\end{tabular}

\vspace{2pt}
\begin{minipage}{0.97\textwidth}\footnotesize
All detectors share $\alpha = 0.02$ and floor $\tau^{\min} = 8 \times 10^{-4}$; bounds are for the binding disturbance window (other windows relax to the floor). Bold: Scenario-D $f_{\mathrm{end}}$ below the 60.5~Hz relay setting. Peak $|\Delta V|$ under Scenario~D: 0.123--0.136~p.u.
{\revb $^{\ddagger}$Bound from the one-step prediction-error threshold. $^{\S}$Separate run with $\beta_{wac}=19$ (benign $\mathcal{J}_B = 172.39$); compare within the row only.}
\end{minipage}
\vspace{-12pt}
\end{table*}

Table~\ref{tab:static_vs_ml_bdd} shows that residual attack severity depends primarily on decision-boundary width rather than detector family. On the 39-bus system, DBSCAN, isolation forest, and one-class SVM learn the same disturbance-window bound and therefore produce numerically identical attacks \rev{that reach 60.5~Hz at 880~TS}. The LSTM learns a wider \rev{frequency} boundary \rev{in the disturbance window} than both these models and the static rule, \rev{which raises the objective by 30.8\%. Its crossing nevertheless occurs later (1000~TS): because~\eqref{eq:attacker-objective} accumulates frequency and voltage deviation over the whole horizon, a larger objective does not imply an earlier crossing. On the 118-bus system, the wider LSTM bound yields both a larger objective (1140.90 vs.\ 1058.93) and an earlier crossing (890 vs.\ 1000--1020~TS) than DBSCAN and one-class SVM, which attain the same objective through alternative optimal trajectories.} Detector sophistication alone, therefore, provides no security benefit unless it yields a tighter, well-calibrated representation of benign controller variation. \rev{No learned bound prevents the crossing; across all eight configurations, it is delayed by 310--600~TS (about 5-10~s).}

Residual voltage excursions remain similar across detectors because~\eqref{eq_frequency_constraints}-~\eqref{eq_wac_constraints} limits the rate of change, not cumulative drift. Restricting rapid frequency manipulation consequently redirects the attack toward the excitation path, motivating a cumulative-displacement monitor in addition to rate-based detection. \rev{Learned constraints change the solve time in both directions (e.g., 49.7 to 89.1~s on the 39-bus and 509.6 to 401.6~s on the 118-bus system), and every solve completes in under 9~min.} Moreover, the 118-bus results preserve the vulnerability and mitigation trends observed on the 39-bus system, indicating that neither disappears as system scale and coupling increase.

%% file: sections/Discussion.tex
\section{Discussion}
\label{sec:discussion}

The results yield \rev{four} system-level findings. First, the wide-area loop reduces benign oscillations by only about 10\%, yet \rev{with wide-area access the optimal attack attains $3.3\times$ and $8.1\times$ the benign objective on the 39- and 118-bus systems}, while more than halving the time to a 0.5~Hz excursion (Section~\ref{sub:impact_accessibility}). Because one wide-area signal reaches the excitation paths of multiple generators, its corruption creates a coordinated perturbation that would otherwise require device-level AGC or AVR access. Security priority should therefore reflect compromise impact rather than benign contribution alone.

Second, learned constraints reduce the objective by \rev{15.5--55.0\%} and \rev{delay the 60.5~Hz over-frequency crossing by about 5--10~s, yet every configuration retains an optimal stealthy attack that still reaches it and whose objective exceeds twice its benign baseline.} Learned detection thus increases attack cost and response time without eliminating the attack surface. Third, residual severity depends more on boundary width than \rev{on} detector family\rev{: on the 39-bus system three distinct detectors learn the same bound and produce identical attacks, whereas the wider LSTM frequency bound permits a 30.8\% larger objective (Table~\ref{tab:static_vs_ml_bdd}).} Learned boundaries should therefore be audited against the static thresholds they replace. \rev{Fourth, voltage excursions of 0.12--0.14~p.u.\ persist across detectors because rate-based bounds restrict rapid changes but not cumulative drift, which motivates} combined rate-of-change and cumulative-displacement monitoring.

\smallskip
\textbf{Limitations.}
The framework models immediate attack impacts but excludes cascading failures and communication latency. Linearized power flow introduces small deviations from the ePHASORSIM AC response (Fig.~\ref{plot:wampac_gurobi}), while exact controller selection is intractable at the 118-bus scale; consequently, the surrogate-based result represents a lower bound on attainable attack severity. Reinforcement learning could improve adaptation and execution speed under nonlinear or time-varying conditions, but does not guarantee optimality or constraint satisfaction; under an identical formulation, the MILP \rev{solved to optimality} upper-bounds any feasible policy. Comparing these approaches and developing policy-guided optimization are left for future work.

%% file: sections/Conclusion.tex
\section{Conclusion}
\label{sec:conclusion}
In this paper, we proposed an attack-resiliency analytics framework for WAMPAC-enabled power grids, in which the optimal stealthy FDI attack is formulated as a mixed-integer linear program over the coupled wide-area, secondary, and primary control dynamics. The anomaly detector is modeled as a replaceable constraint set, enabling static BDD rules and learned detection boundaries to be evaluated within a unified formulation. The results on the IEEE 39- and 118-bus systems demonstrate that access to the wide-area channel substantially amplifies attack severity and accelerates relay-relevant frequency excursions. Although learned boundaries reduce attack severity and delay over-frequency crossings, they do not prevent them, and their effectiveness depends on boundary tightness rather than detector sophistication. These findings indicate that WAMPAC channels should be prioritized according to their compromise impact and that learned detectors should be audited against the static thresholds they replace. Future work will extend the framework to incorporate cascading failures, communication latency, and cumulative-displacement monitoring, and will evaluate its scalability on larger systems.


\section*{Acknowledgement}
This work is supported by the Department of Energy (DOE) under Award\# DE-CR0000024. Any opinions, findings, conclusions, or recommendations expressed in this material are those of the authors and do not necessarily reflect the
views of the DOE.